\documentclass[twocol]{ametsocV6.1}

\usepackage{amsmath}
\usepackage{amssymb}
\usepackage{graphicx}
\usepackage{overpic}
\usepackage{xcolor}
\usepackage{hyperref}

\usepackage{booktabs}
\usepackage{multirow}
\usepackage{siunitx}
\usepackage{tabularx}

\usepackage{oplotsymbl}
\usepackage{subcaption}

\usepackage{textcomp}     

\usepackage{tikz}
\newcommand{\fatx}{
  \begin{tikzpicture}[baseline=-0.5ex,scale=0.3]
    \draw[line width=2.5pt] (-0.3,-0.3) -- (0.3,0.3);
    \draw[line width=2.5pt] (0.3,-0.3) -- (-0.3,0.3);
  \end{tikzpicture}
}

\newcommand{\fatplus}{
  \begin{tikzpicture}[baseline=-0.5ex, scale=0.3]
    \draw[line width=2.5pt] (0,-0.3) -- (0,0.3);   
    \draw[line width=2.5pt] (-0.3,0) -- (0.3,0);   
  \end{tikzpicture}
}

\hypersetup{
	colorlinks=true, 
	linkcolor=blue,  
	citecolor=blue,  
	urlcolor=blue,   
	pdfborder={0 0 1} 
}

\newcommand\bV{{\mathbf{V}}}

\newcommand\bFF{{\mathbf{GS}}}
\newcommand\bGG{{\mathbf{AR}}}
\newcommand\bHH{{\mathbf{AP}}}
\newcommand\cM{{\mathcal M}}
\newcommand\cH{{\mathcal H}}
\newcommand\bM{{\mathbf{M}}}
\newcommand\hV{{\widehat{V}}}
\newcommand\hI{{\widehat{I}}}
\newcommand\hr{{\widehat{\rho}}}
\newcommand\hbV{{\widehat{\bV}}}
\newcommand\bC{{\mathbf{C}}}
\newcommand\bD{{\mathbf{D}}}

\newcommand\bX{{\mathbf{X}}}
\newcommand\bZ{{\mathbf{Z}}}

\newcommand\bH{{\mathbf{H}}}

\newcommand\bT{{\mathbf{\Theta}}}
\newcommand\bP{{\mathbf{\Pi}}}

\title{Local wind speed forecasting at short time horizons based on Numerical Weather Prediction and observations from surrounding stations}

\authors{Roberta Baggio,\aff{a} 
	Killian Pujol,\aff{b}
	Florian Pantillon,\aff{b} 
	Dominique Lambert,\aff{b} 
	Jean-Baptiste Filippi,\aff{a} 
	Jean-François Muzy,\aff{a} \correspondingauthor{J.F. Muzy, muzy@univ-corse.fr} 
}

\affiliation{\aff{a}{Laboratoire Sciences Pour L’Environnement (SPE), UMR 6134,  CNRS, Université de Corse} \\
\aff{b}{Laboratoire d’Aérologie (LAERO), Université de Toulouse, CNRS, IRD}}

\date{\today}

\abstract{
This study presents a hybrid neural network model for short-term (1-6 hours ahead) surface wind speed forecasting, combining Numerical Weather Prediction (NWP) with observational data from ground weather stations. It relies on the MeteoNet dataset, which includes data from global (ARPEGE) and regional (AROME) NWP models of the French weather service and meteorological observations from ground stations in the French Mediterranean. The proposed neural network architecture integrates station observations from the last few hours and AROME and ARPEGE predictions on a small subgrid around the target location. The model is designed to provide both deterministic and probabilistic forecasts, with the latter predicting the parameters of a suitable probability distribution that notably allows us to capture extreme wind events.
Our results demonstrate that the hybrid model significantly outperforms baseline methods, including raw NWP predictions, persistence models, and linear regression, across all forecast horizons. For instance, the model reduces RMSE by up to 30\% compared to AROME predictions. Probabilistic forecasting further enhances performance, particularly for extreme quantiles, by estimating conditional quantiles rather than relying solely on the conditional mean. Fine-tuning the model for specific stations, such as those in the Mediterranean island of Corsica, further improves forecasting accuracy.
Our study highlights the importance of integrating multiple data sources and probabilistic approaches to improve short-term wind speed forecasting. 
It defines an effective approach, even in a complex terrain like Corsica where localized wind variations are significant.}

\begin{document}
	

\maketitle

\section{Introduction}
Wind speed forecasting is an important issue that concerns many applications, from public safety to renewable energy production. In the energy sector, where the production of wind farms is inherently variable, accurate wind forecasts over short-term horizons (few hours) are crucial for enabling grid operators and wind farm managers to balance supply and demand, plan operations effectively, and mitigate risks \citep{shaw_second_2019}. Likewise, accurate short-term wind speed forecasts are critical for safeguarding public welfare by supplying essential information for aviation, maritime operations, and other weather-sensitive activities, especially in the case of severe wind episodes \citep{pinto_atmospheric_2019}. 
However, the turbulent and intermittent nature of wind, combined with rapid variations in local conditions, makes short-term wind prediction a highly complex task \citep{pantillon_formation_2020}. 

Traditional weather forecasting relies heavily on Numerical Weather Prediction (NWP) models, which simulate atmospheric dynamics based on physical principles and use initial conditions derived from observational data. National weather services usually combine NWP models at the global and regional scales to produce forecasts. For instance, Météo France uses the ARPEGE and AROME models:
ARPEGE \citep{courtier_strategy_1994}, a global model with a refined resolution of about 5 km over Europe and France, provides broader-scale forecasts up to 5~days ahead but is limited in its ability to capture local phenomena;
AROME \citep{seity_arome-france_2011} is a high-resolution limited-area model 
downscaled from ARPEGE,
specifically designed for 
shorter-range forecasts up to 2~days ahead and
with a horizontal resolution of 1.3 km, which makes it effective in resolving weather phenomena at smaller scales and deep convection in particular.
Despite the considerable progress made in recent decades \citep{bauer2015}, NWPs are computationally intensive and often struggle to provide accurate predictions at fine spatial and temporal scales. 
Designed primarily for meso to synoptic scales $\mathcal{O}$(10--1000~km), 
NWP models are less effective in capturing localized variations and rapidly evolving phenomena such as wind gusts \citep{pantillon_forecasting_2018} or local convective episodes \citep{yano_scientific_2018}.
Moreover, since the atmosphere is a chaotic system, it is desirable to provide forecasts with an estimate of the associated uncertainty, that is, to make probabilistic predictions. For this reason, operational centers produce ensemble forecasts, which consist of multiple runs of the same NWP, obtained using stochastically perturbed initial conditions and physical equations, or by aggregating results from different models (or both; \citet{swinbank_tigge_2016}), a fact that makes NWP forecasts even more costly.

To address those limitations of NWP models for wind speed ``nowcasting'' (namely, forecasting at horizons up to few hours) a range of statistical, Machine Learning (ML) and Deep Learning (DL) methods has emerged. These methods mainly rely on statistical inference or time-series model based approaches that use historical data and focus on past observed patterns to forecast future observations (see, e.g., \citet{WS_Review_2014} and references therein). Some of these techniques involve the development of specific stochastic models in order to account for observed fluctuations, particularly within the realm of time series analysis \citep{BaileMuzyPoggi2011}. 
Most of the recent approaches leverage more advanced ML and DL techniques. In particular, artificial Neural Networks (NN) have proven to be very efficient in capturing the nonlinear dependencies that define wind dynamics. Numerous neural network models with many different architectures have been proposed so far. The most commonly used models are Recurrent Neural Networks, 
temporal or spatial Convolutional Networks, 
Graph Convolutional Networks 
or transformer based models. 
A comprehensive overview of this rapidly evolving field can be found in numerous review papers (see, e.g., \citet{WS_Review_2020,WS_Review_2021}).

Besides pure NWP or pure statistical approaches, hybrid methods that combine the two have been experimented in order to produce better predictions, especially at short-term \citep{Huang12}.
In this paper, we define hybrid models as machine learning-based architectures trained on both NWP forecasts and other source of data (like observational data), where the model structure explicitly allows for the fusion of these inputs—potentially through separate branches or feature encoders. These models may incorporate NWP output as predictors within statistical frameworks, such as time series models or ML approaches (like e.g., NN models). One expects that the NWP forecasts provide valuable information about large-scale atmospheric conditions, while processing observational data allows one to account for local effects, model biases, and unresolved sub-grid processes to improve wind speed or wind power nowcasting \citep{WS_NWP_ML_2018,WS_NWP_ML_2021,hybridWPforecast21, WS_NWP_ML_2022}.

It is noteworthy that such approaches should not be confused with the statistical post-processing of NWP forecasts \citep{vannitsem21}.
Indeed, it is well known that direct forecasting from NWP, regardless of whether it is global or regional, often exhibits systematic biases and limitations, particularly at the local scale and close to the surface for variables like wind speed \citep{Zamo_2016}. These errors stem from factors such as insufficient model resolution, incomplete representation of physical processes, and inaccuracies in initial conditions. For that reason, operational centers consider various post-processing methods in order to remove those biases and improve prediction. The most commonly used method is the “Model Output Statistics” (MOS), which consists of regressing observed data from raw NWP outputs. This framework has been extended to probabilistic forecasting with the “Ensemble Model Output Statistic” (EMOS), designed to handle ensemble predictions \citep{Gneiting05}.
However, MOS and EMOS typically rely on linear statistical corrections and do not incorporate other data, e.g., observational time series, as separate input features. More elaborated statistical methods, that can capture non-linear, complex relationships between NWP outputs and observed weather variables—like Kalman filtering \citep{KalFil17}, random forests \citep{Zamo_2016}, or NN models \citep{rasp2018,Vel21}—have been proposed more recently. Consequently, advancements in NWP post-processing methods and hybrid model approaches have made them increasingly difficult to distinguish.
Still, the key distinction we adopt in this work is that hybrid models aim to jointly learn from various types of forecasts and observations within an integrated learning framework, whereas post-processing focuses on correcting NWP forecasts—often using simpler statistical models.

This study contributes to the development of hybrid statistical methods for reliable nowcasting of surface wind speed. It leverages the MeteoNet database, an open-access resource provided by Météo France, which includes both NWP and observational data  \citep{meteonet_dataset}.
MeteoNet covers two regions of France (the Southeast and Northwest) and provides essential meteorological variables at hourly and sub-hourly temporal resolutions, making it well-suited for training and validating ML models for short-term, location-specific forecasts. We focus on the French Mediterranean region, particularly Corsica, a French Mediterranean island (80 × 180 km) characterized by complex mountainous terrain and frequent high-impact meteorological events such as windstorms, heavy rainfall, thunderstorms and Saharan dust incursions \citep{scheffknecht_climatology_2017,coquillat_saetta_2019,lfarh_downward_2023}. The Mediterranean region is also a recognized climate change hotspot, motivating various research initiatives (e.g., \citet{drobinski_hymex_2014}, \citet{hatzaki_medcyclones_2023}).
We introduce a NN model to enhance short-term wind speed forecasting in this geographically complex setting. Our model integrates NWP data from both global (ARPEGE) and regional (AROME) models with station observations to generate deterministic and probabilistic forecasts for lead times of one to six hours at any given selected station.
By combining the strengths of physically based NWP models and data-driven ML techniques, our proposed framework aims at capturing both large-scale weather patterns and localized wind variations in order to improve forecasting accuracy and adaptability to site-specific conditions.
Our framework is designed to support operational needs in the wind energy sector, providing actionable insights for planning, risk mitigation, and decision-making.

Finally, it is relevant to mention that a similar approach utilizing the same dataset was recently proposed by \citet{Marcille24} to forecast wind speed at offshore locations where observational data is unavailable.

The paper is organized as follows. Section~\ref{sec:data} introduces the MeteoNet dataset used throughout this study. Section~\ref{sec:prediction} defines the forecasting task under consideration, along with the associated evaluation framework. Section~\ref{sec:models} presents the different forecasting models and baselines included in our comparison. In particular, we introduce two hybrid architectures, denoted by $\mathcal{M}$ and $\mathcal{M}'$, which combine NWP outputs with weather station data from MeteoNet to produce deterministic and probabilistic short-term forecasts of near-surface wind speed, respectively.
Section~\ref{sec:Results} reports the empirical results. Subsection~\ref{sec:SE} provides a comparative performance analysis of $\mathcal{M}$ and $\mathcal{M}'$ against several baselines across all stations.
Subsection~\ref{sec:Corsica} focuses on models specifically fine-tuned for stations located in Corsica, where we evaluate both standard point forecasting (based on error minimization) and the ability to predict extreme wind conditions through threshold exceedance events. Section \ref{sec:discussion} presents limitations of the proposed approach, identifies potential improvements and considers some operational aspects.  Finally, Section~\ref{sec:Conclusion} summarizes our main findings and suggests directions for future work.

\section{Data}
\label{sec:data}
\subsection{The MeteoNet database}
The results presented in this article are based on the MeteoNet database, which is
a comprehensive and freely available meteorological dataset published by Météo France, the French national meteorological service \citep{meteonet_dataset}. It has been specifically designed to support research and development in meteorology, climatology, and machine learning applications. The MeteoNet database contains high-resolution meteorological data over, respectively, the Northwest and Southeast quarters of the France territory covering the period from 1 January 2016 to 31 December 2018. It includes observations from ground-based weather stations, gridded forecasts from NWP models, and auxiliary geospatial information.

\begin{figure*}[h!]
	\centering
	\includegraphics[width=1\linewidth]{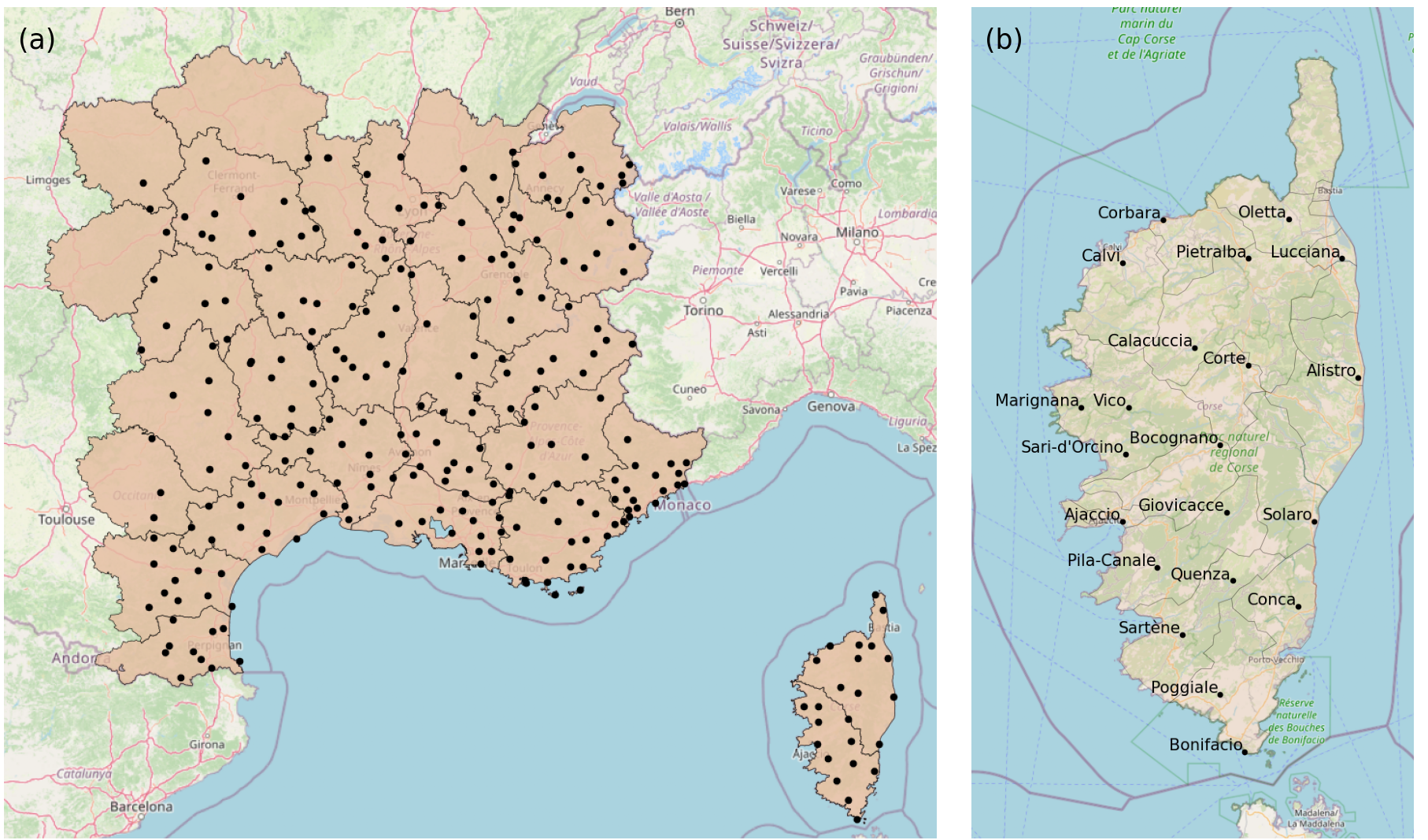}
	\vspace*{-0.5cm}
	\caption{(a) Geographical extent of the MeteoNet Southeast database, with the 278 ground station localizations ($\bullet$). (b) Location and names of the 21 target ground stations in Corsica. }
	\label{fig:sta_loc}
\end{figure*}

In this work, we will consider exclusively the Southeast data, influenced mostly by a Mediterranean climate with, as mentioned in the introduction, a special focus on sites spread all over Corsica. 
Observation data are provided by the Météo France ground station network. 
A subset of 278 stations was selected by setting a data availability threshold to ensure statistical significance.
The position of each of these selected stations is shown in 
Fig.~\ref{fig:sta_loc}(a). 
They provide, at a time resolution of 6 min, time records of different weather variables as notably 10 $m$ wind speed and direction and 2 $m$ air temperature.
In this study, we also use the forecast data from M\'et\'eo France operational NWP models ARPEGE and AROME included in the MeteoNet database with 0.1$^{\circ}$ and 0.025$^{\circ}$ horizontal resolutions, respectively.
A single run is provided for each day, the one beginning at 00 UTC that spans the forecast ranges from 00 to 24 UTC, with a time step of 1~h for the AROME data and 1~h or 3~h for the ARPEGE data, depending on the time horizon. 
AROME data are composed of different surface fields (notably $10$ m wind speed, $2$ m temperature, relative humidity and mean-sea-level pressure) whereas ARPEGE provides predictions of various fields (like horizontal wind speed components or temperature) at 7 different isobar levels spanning the low to mid troposphere
and at 7 different heights in the lower troposphere for the pressure variable.

To prepare the dataset for use in the training, validation, and test phases of the models presented in Sec. \ref{sec:models}, we implement a two-step data preprocessing workflow. In the first step, the raw MeteoNet Southeast dataset is reorganized into a station-centric format. Specifically, for each available station and each day, we generate a NetCDF file that compiles all relevant data: the station’s ground-based time series for the day (including context from adjacent days), as well as AROME and ARPEGE model outputs in the form of spatial sub-images centered on the station. This stage is purely structural, and no filtering or cleaning is applied, in order to preserve the integrity of the original data. The second step, which is closely tied to the specific prediction task defined in Sec. \ref{sec:prediction}, involves transforming these NetCDF files into model-ready inputs. Each file is read and processed to construct the input tensors and corresponding output values used during training, validation, and testing (see Sec. \ref{sec:model_inputs} for details on the input structure). During this stage, we discard any samples (i.e., specific station-day pairs) containing missing or invalid values, normalize the input features, and store the resulting data in a unified format—either as a large binary file or an in-memory arrays —compatible with TensorFlow or PyTorch data generators. This two-stage process ensures flexibility, reproducibility, and robust quality control across the entire data pipeline.

\subsection{Wind speed station data in Corsica}
\label{sec:corsica_stations}

Some of our results detailed in the following specifically focus on ground stations in Corsica. These stations  correspond to the subset of locations represented in 
Fig.~\ref{fig:sta_loc}(b). 

For comparison purpose, we also utilize a specific dataset of wind speed and direction time series previously used in \citet{BaggioMuzy2024}. This dataset comprises hourly wind speed measurements recorded by Météo France at various locations across Corsica from 1 January 2011 to 31 December 2020. It can
be used to train a neural network model such as the one 
introduced in \citet{BaileMuzy2023} which relies exclusively on past observed data from the target station and its 15 closest neighboring stations.
In section \ref{sec:Corsica_ft},
we compare  a model relying exclusively on observed data to the hybrid approach advocated in this paper on four representative sites in Corsica, namely Ajaccio, Lucciana, Renno, and Figari. These four selected sites together with their neighboring stations have a low proportion of missing data, with less than 15\% of the total possible data points unavailable. 
For this analysis, the period from 1 January 2011 to 31 December 2015 is used as the training sample for model of \citet{BaileMuzy2023}, while the comparison is performed on the same validation and test subsets we define for the MeteoNet dataset over the 2016–2018 period (see Sec. \ref{sec:subsets}).

%
%
%
%

\section{Prediction task at multiple horizons}
\label{sec:prediction}
In this section, we establish the primary notations and formally define the forecasting problem addressed in this study. Our main objective is to predict the surface wind speed at a specific site over short-term forecast horizons, typically ranging from 1 to 6 hours ahead. As explained in the introduction, a central motivation for this task,  beyond producing accurate wind speed forecasts, is to enable the assessment of specific events such as whether the wind speed will exceed a predefined extreme value threshold—information that is critical for decision-making in weather-sensitive applications. To model this predictive task, we consider two alternative strategies. The first is a deterministic approach, which yields directly a point forecast—i.e., a single predicted value—for future wind speed. The second is a probabilistic approach, which estimates the full conditional distribution of future wind speeds, thereby supporting both point predictions and probabilistic quantification of events such as threshold exceedance. We evaluate and compare the performance of these approaches using appropriate metrics.

Let $V^s_{d,t}$ be the surface wind velocity (in $m \; s^{-1}$) at time $t$ of day $d$ at a given fixed location $s$. In all our experiments we consider the hourly frequency and thus, for a given day $d$, $t$ is expressed in hours. The upper character $s$ can represent the longitude/latitude coordinates of the target location or simply the site name. Notice that $V^s_{d,t}$ is directly recorded if $s$ corresponds to a site where a ground station is installed. 
The time $t$ of the day when the prediction is performed will be called the {\em initial time}.
Let 
$
 \bH = (h_1,h_2, \ldots, h_{N_H})
$
be a set of $N_H$ fixed time horizons values such that $h_1 < h_2 < \ldots < h_{N_H}$. For the sake of simplicity and to avoid handling horizons too far away from NWP run initial time, we restrict ourselves to the set of next $N_H = 6$ upcoming hours, namely $\bH  = (1,2,3,4,5,6)$ hours.  
We denote by $\bV^s_{d,t+\bH}$ the $6$-dimensional vector of future velocities at multiple horizons:
$$
\bV^s_{d,t+\bH} = \begin{pmatrix}
	V^s_{d,t+1} \\
	V^s_{d,t+2} \\
	\vdots \\
	V^s_{d,t+6}
\end{pmatrix} \; .
$$

In the literature a single horizon time $h_i$ is also often referred to as the {\em lead time} (e.g. in \citet{Marcille24}). The forecasting task we consider is therefore to compute, 
at each $d$ and $t$, for each horizon $h_i \in \bH$, a velocity value $\widehat{V}^s_{d,t+h_i}$ that estimates ``at best'' $V^s_{d,t+h_i}$. 
The corresponding 6-dimensional prediction vector will be denoted as
$\hbV^s_{d,t+\bH}$. Notice that for a single horizon \( h \) and initial time \( t \), the time \( t + h \) is often referred to as the \emph{valid time} or the \emph{target time}.

At this stage, it is important to mention that our goal is not to compare models relying on NWP in general, which may be initialized multiple times per day, but specifically those from the 00~UTC run. Indeed, the MeteoNet dataset only includes NWP forecasts initialized at 00~UTC (hereafter referred to as the \textit{initialization time}), a forecast made at time~$t$ with horizon~$h$ corresponds therfore to an ``\textit{effective horizon}'' of~$t + h$. In the following, we evaluate performance at fixed forecast horizons~$h$, which is more relevant operationally than fixing the effective time~$t + h$. This means we average results over all initial times~$t$, at fixed~$h$, thereby covering weather events occurring at different times of day. In this framework, the performance of models that involve only AROME or ARPEGE predictions effectively integrates a range of effective horizons, which depends only marginally on $h$ for t large enough. As a result, we expect NWP-based results to exhibit very limited dependence on $h$. 

The most commonly used definition of  the ``best prediction"
$\hbV^s_{d,t+\bH}$ is the one that minimizes the Mean Squared Error (MSE):
\begin{equation}
	\mathcal{E}^{mse} \!\!  = {\mathbb E} \left( \left\|\bV^s_{d,t+\bH}-\hbV^s_{d,t+\bH} \right\|^2 \right)
	= \sum_{i=1}^6 {\mathbb E} \left( \left|\hV^s_{d,t+h_i}-V^s_{d,t+h_i} \right|^2  \right)
\end{equation}
where $\mathbb E$ represents the mathematical expectation 
and $\left\|\bZ\right\|^2$ is the squared euclidian norm of $\bZ$.
Along the same line, one defines the conditional mean squared error as:
\begin{eqnarray}
	\nonumber
	\mathcal{E}^{mse}_{d,t}&  = & {\mathbb E} \left( \left\|\bV^s_{d,t+\bH}-\hbV^s_{d,t+\bH} \right\|^2 \Big| \; {\cal F}_{d,t} \right) \\
	  & =&  \sum_{i=1}^6 {\mathbb E} \left( \left|\hV^s_{d,t+h_i}-V^s_{d,t+h_i} \right|^2 \Big| \; {\cal F}_{d,t} \right)
\end{eqnarray}
where ${\cal F}_{d,t}$ represents all the ``information'' available at day $d$, time $t$ and $\mathbb{E}(. |  {\cal F}_{d,t} )$ stands for the mathematical expectation conditionnally to ${\cal F}_{d,t}$ meaning the expectation as respect to the conditional probability law of future outcome of weather conditions at times $t+h_i$, $h_i \in \bH$.
In practice, at each day $d$ and time $t$, the abstract information set $\mathcal{F}_{d,t}$ available at site $s$ is encoded into a high-dimensional feature vector $\bX^s_{d,t}$. This vector may include, for instance, weather observations at station $s$ and its neighboring stations, the day of the year, and other relevant variables (see next section for details). Accordingly, the conditional expectation $\mathbb{E}(\cdot \mid \mathcal{F}_{d,t})$ is approximated by $\mathbb{E}(\cdot \mid \bX^s_{d,t})$.

From the law of iterated expectations, we simply have
$$
\mathcal{E}^{mse} = {\mathbb E} \left(	\mathcal{E}^{mse}_{d,t}  \right)
$$
and therefore minimizing at all $t,d$, $\mathcal{E}^{mse}_{d,t}$ leads to minimizing
the global error $\mathcal{E}^{mse}$.

It is well known that the vector $\bM^s_{d,t+\bH}$ that minimizes $\mathcal{E}^{mse}_{d,t}$ is simply {\em the conditional mean}:
$$
\bM^s_{d,t+\bH} = \mathbb {E} \left(\bV^s_{t+\bH} \Big| \; \bX^{s}_{d,t}  \right) \; .
$$

In that respect, for any station $s$, at any time $t$ and day $d$, minimizing the MSE is equivalent to building $\hbV^s_{d,t+\bH}$ as the best approximation of the conditional expectation $\bM^s_{d,t+\bH}$ of the multi-horizon velocity vector $\hbV^s_{d,t+\bH}$. 

As mentioned above, we will consider two distinct methods to do that, a deterministic approach and a probalistic appraoch.

\subsection{Deterministic approach}
A commonly used possibility to estimate $\bM^s_{d,t+\bH}$ is to seek for a {\em model}  $\cM(\bX^s_{d,t}, \bT)$ that directly outputs $\bV^s_{d,t+h}$:
\begin{equation}
	\label{eq:mod_gen}
	\hbV^s_{d,t+\bH} = \cM(\bX^s_{d,t}, \bT)
\end{equation}
where $\bT$ is a set of model parameters which is chosen in order to minimize 
$\mathcal{E}^{mse}$.  This latter can be estimated empirically as the following loss function:
\begin{eqnarray}
 \nonumber
	\mathcal{L}^{mse}   & = &   N^{-1} \sum_{k=1}^N  \left\|\bV^s_{d,t_k+\bH}-\hbV^s_{d,t_k+\bH} \right\|^2   \\ 	\label{eq:mse_loss}
	& =   & 
	N^{-1} \sum_{k=1}^N \sum_{i=1}^{6} \left(\hV^{s_k}_{d_k,t_k+h_i} - V^{s_k}_{d_k,t_k+h_i} \right)^2
\end{eqnarray}
where $N$ is the number of available samples (in the training subset, see below) which corresponds to a fixed fraction of the total number of different triplets 
``station'', ``day'' and ``initial time'' ($s$,$d$,$t$) for which 
the observed velocity vector $\bV^s_{d,t+\bH}$ and the input features $\bX^s_{d,t}$ are available. 

To assess the forecasting performance of a model, the most commonly used score is the Root Mean Square Error (RMSE), defined as:
\begin{equation}
	\label{eq:rmse}
	\text{RMSE}   =   \sqrt{ 	\mathcal{L}^{mse}  }
\end{equation} 
where $	\mathcal{L}^{mse} $ is computed on either the validation or test subset.
RMSE corresponds to the empirical deviation (in $m\; s^{-1}$) of observed velocity as respect to the prediction $\hbV$.

\subsection{Probabilistic approach}
 Alternatively to the previous deterministic  (also called ``single-point'') forecast, 
 one can estimate the full conditional probability distribution from which
 the conditional mean is then derived (see Eqs. \eqref{eq:mod_gen_p}, \eqref{eq:c_mean_p}). 
 Probabilistic learning methods has become increasingly important in recent years as it captures not only the expected outcome but also uncertainties around this expectation and therefore provides notably insights into the likelihood of extreme events.  They can rely on different approaches, like 
 quantile-based methods which consists in directly estimating specific quantiles as the median or the 90th percentile, without assuming 
 any particular probability distribution (see, e.g., \citet{koenker_2005_quantile,park_2022}). 
 Another  widely used strategy involves coupling a sequential model with a likelihood model, where the latter governs how observations are generated from underlying hidden or latent states (see, e.g., \citet{GP,DeepSS_2018}).
 The method we use in this work relies on such approach and is inspired by \citet{DeepAR}. It consists in
 estimating, at each time step, the set of parameters characterizing the conditional law of upcoming velocity chosen to belong to a given parametric class of distributions. 
 This approach is very parsimonious, offering a complete characterization of the forecast distribution, and it can easily integrate additional information, such as prior distribution, using within the framework of Bayesian inference. 
 More specifically, for each station $s$, day $d$ and initial time step $t$, the tensor of parameters $\bP_{s,d,t}$ characterizing, $\rho(\bV ; \bP_{s,d,t})$, the conditional law of $\bV^s_{d,t+\bH}$ as the output of a neural-network model $\cM'$:
 \begin{equation}
 \label{eq:mod_gen_p}
 \bP_{s,d,t} = \cM'(\bX^s_{d,t}, \bT) \; 
 \end{equation}
 where again, $\bT$ stands for the set of parameters characterizing the model $\cM'$ that have to be learned. 
 For that purpose we suppose that the multivariate density is simply the product of 
 monovariate densities:
 $$
  \rho(\bV ; \bP_{s,d,t}) = \prod_{i=1}^6 \rho(V_i;\bP^i_{s,d,t})
 $$
 where $\bP^i_{s,d,t}$ is the vector parameters of the conditional
 law of $V^s_{d,t+h_i}$. The model $\cM'$ parameters are learnt 
 by minimizing the negative log-likelihood  (also called  \emph{logarithmic score}): 
 \begin{equation}
 	\label{def:logS_multi}
 		\mathcal{L}^{\ell}   = - N^{-1} \sum_{k=1}^N \sum_{i=1}^6  \ln \rho(V^{s_k}_{d_k,t_k+h_i} ; \bP^i_{t_k}) \; .
 \end{equation}
 
 In this framework, the conditional mean estimation $\hbV_{t+\bH}^s$ is computed as follows:
 \begin{equation}
 	\label{eq:c_mean_p}
 	\hV^s_{d,t+h_i}  =  \int_{{\mathbb R}} z \rho(z ; \bP^i_{s,d,t}) \;  dz \; . 
 \end{equation}
 
We choose to base our computations on $\rho(z)$ belonging to the class of so-called ``Multifractal-Rice'' (M-Rice) distribution, which has proven to be particularly suitable for wind speed forecasting  when compared to other distributions like Weibull or Gamma \citep{BaggioMuzy2024}. 
The M-Rice distribution, introduced in \citet{BaMuPo11}, is motivated by the random cascade framework used in fully developed turbulence. It extends the classical Rice distribution by allowing its scale parameter to be random (typically modeled as a log-normal variable). The resulting distribution is governed by three parameters: \((\nu, \sigma^2, \lambda^2)\). Here, \(\nu\) represents a mean parameter, \(\sigma^2\) a dispersion parameter, and \(\lambda^2\), often referred to as the \emph{intermittency parameter}, controls the shape of the distribution. In particular, larger values of \(\lambda^2\) lead to \emph{heavier tails}, capturing the increased probability of extreme values. The mathematical definition of M-Rice distribution, the interpretation of its parameters and details for numerical estimations are provided in \ref{app:M-Rice}. 

\subsection{Extreme event prediction and associated performance scores}
 \label{sec:ext_metrics}
 As already emphasized, since our goal is to compare models providing single-point predictions,
 we treat model $\cM'$ of previous section that provides a probabilistic forecasting as an intermediate step for generating deterministic predictions for example through Eq. \eqref{eq:c_mean_p} which performance can be evaluated using the RMSE metrics defined in \eqref{eq:rmse}.

To evaluate a model’s ability to forecast \emph{extreme wind events}, we focus on \emph{threshold exceedances}. The occurrence of a wind with amplitude greater than some threshold $Q>0$ is represented by a binary indicator variable defined as follows:
Let  ${\cal H}$ be the Heaviside function and, for any site \( s \), at time \( t \) on day \( d \), the binary variable \( I_{s,d,t}(Q) \) is defined as  
\begin{equation}
	\label{eq:def_Iq}
	I_{s,d,t}(Q) = \cH( V^s_{d,t}-Q) = 
	\begin{cases} 
		1 & \text{if } V^s_{d,t} \geq Q \\ 
		0 & \text{otherwise}.
	\end{cases}
\end{equation}
Within this context, predicting strong wind events at a given location amounts to forecasting the occurrence of threshold exceedances with some threshold corresponding to a high quantile relative to the site's wind climatology.  
Let $Q_C^{s}(p)$ denote such a quantile of probability level $p$ close to $p=1$. If  the climatological distribution at station $s$ is denoted as $\rho^s_C(V)$, one has, by definition:
\begin{equation}
	\label{eq:def_quantile}
	\int_0^{Q_C^{s}(p)} \rho^s_C(V) \; dV = p  \; .
\end{equation}
In that respect, if $1-p \ll 1$, $I_{s,d,t}(Q_C^s(p))$  indicates the occurrence of strong wind speeds greater than a threshold in the tail of the site's climatological distribution (i.e., corresponding to a high level quantile).
Predicting strong events occurrence thus amounts to predict the value of $I_{s,d,t+h}(Q^s_C(p))$ at each time $t$, day $d$, for each station $s$ and horizon $h$.

Assessing the quality of forecasts in this context thus amounts to evaluating the accuracy of binary predictions  which is  a \emph{classification task}—in contrast to the \emph{regression setting} considered earlier. In this classification framework, various \emph{skill scores} can be used to evaluate performance, depending on the aspect of predictive quality one wishes to emphasize (e.g., sensitivity, specificity, overall accuracy). For a comprehensive overview of relevant metrics and their properties, we refer the reader to \citet{Jolliffe2004} and \citet{WILKS2019369}. Among all possible scores, the Peirce Skill Score (PSS) is a widely used metric for evaluating the performance of a binary classification model, particularly in the context of severe weather prediction. It is defined as:  
\begin{equation}
	\label{eq:PSS}
	\text{PSS} = \text{HR} - \text{FA} 
\end{equation}
where HR stands for ``Hit Rate'' and FA for ``False Alarm Rate''. These rates are derived from the contingency table of predictions, where the $N$ forecast/observation pairs $(\hI^s_{d,t}, I^s_{d,t})$ are divided in true positives TP, false positives FP, false negatives FN and true negatives TN (with TP+FP+FN+TN=N). 
More specifically, $\text{HR} =  \frac{\text{TP}}{\text{TP}+\text{FN}}$  corresponds to the fraction of correctly predicted positive events,
while $\text{FA} =  \frac{\text{FP}}{\text{TN}+\text{FP}}$ is the fraction of wrongly predicted, i.e. predicted as positive, negative events.  Unlike accuracy-based scores, PSS accounts for both missed events (false negatives) and false alarms (false positives), providing a balanced assessment of quantifying model performance. Moreover, this score is particularly suitable for forecasting extremes as it weights more strongly correct yes forecasts of rare events, thus discouraging artificial distortions towards the more common class. The PSS ranges from \(-1\) to \(1\), with $\text{PSS}=1$ corresponding to a perfect forecast and $\text{PSS} = 0$ corresponding to unskilled, i.e., purely random, forecast. This means that a negative PSS value reflects prediction performance that is  worse-than-random.

Another score commonly used  when forecasting rare events \citep{Jolliffe2004} is the Critical Success Index (CSI), also known as Threat Score. It is defined as:
\begin{equation}
	\label{eq:CSI}
	\text{CSI} = \frac{\text{TP}}{\text{TP}+\text{FP}+\text{FN}}\,,
\end{equation}
corresponding to the ratio between the times $I^s_{d,t}=1$ has been forecasted correctly and the total number of times the event either happened, has been forecasted or both. The maximum value for CSI is 1, while the lowest possible value is 0. As TN are not considered in Eq. \eqref{eq:CSI}, this score is not artificially inflated by correct forecasts of the more common "no" event. However, contrary to PSS, it is not an equitable score,  which makes comparisons between binary events with different frequency of occurrence not straightforward.  In the following, we refer mostly to the PSS when assessing performance, however, we will show some results in terms of CSI as well. 

In the case of a single point prediction at site $s$, day $d$ and time $t$, whatever the considered score, a natural predictor of $I_{s,d,t+h_i}(Q_C^s(p))$ is simply:
\begin{equation}
	\label{eq:pred_mean}
	\hI^{1}_{s,d,t+h_i} =\cH \Big( \hV^s_{d,t+h_i}-Q_C^s(p) \Big)  \; 
\end{equation} 
where $\hV_{d,t+h_i}^s$ is i-th component of $\hbV_{d,t+\bH}^s$, the output of the considered model ($\cM$ model  or the conditional mean associated with the law predicted by $\cM'$ a probabilistic model). On the other hand, for any station $s$, the probabilistic approach provides,  for each day $d$, initial time $t$ and horizon $h_i$, an estimate of $\hr^s_{d,t+h_i}(z)$, the conditional probability distribution of $V_{t+h_i}^s$ from which one can infer the conditional level $p$ quantile, ${\widehat Q}^s_{t+h_i}(p)$ that satisfies:
\begin{equation}
	\label{eq:def_cond_quantile}
	\int_0^{\widehat{Q}^{s}_{t+h_i}(p)} \hr^s_{d,t+h_i}(z) \; dz = p  \; .
\end{equation}
In this case, 
as shown in \citet{Mason1979} (see also \citet{Jolliffe2004}), one can explicitely build the ``best'' predictor of $I^s_{d,t+h}$, in the sense it is the one that maximizes the PSS, as:
\begin{equation}
	\label{eq:pred_quant}
	\hI^{2}_{s,d,t+h} =\cH \Big( \widehat{Q}^{s}_{t+h_i}(p)-Q_C^s(p) \Big) \; .
\end{equation}
Such a prediction is different from the previous one in Eq. \eqref{eq:pred_mean}
since the expected conditional mean value estimation $\hV^s_{t+h}$ has been replaced by the level-$p$ conditional quantile estimation. 

In the case one wants to optimize the CSI score, as advocated in \citet{Jolliffe2004},
another a more efficient predictor can be built since in this case the optimal probability threshold in Eq. \eqref{eq:def_cond_quantile} is no longer $p$ as for PSS but 
is $p'$ such that $p' = \frac{CSI}{1+CSI}$, which corresponds to $p' \approx 0.33$ for an expected CSI around $0.5$. This leads to the near optimal predictor for CSI metrics:
 \begin{equation}
 	\label{eq:pred_quant_CSI}
 	\hI^{3}_{s,d,t+h}  =\cH \Big( \widehat{Q}^{s}_{t+h_i}(0.33)-Q_C^s(p) \Big) \; .
 \end{equation}
 
 In section \ref{sec:ext_winds}, all these 3 predictors are compared for extreme wind threshold in Corsica.

\section{Neural network and baseline models}
\label{sec:models}
\subsection{Model inputs}
\label{sec:model_inputs}

The objective of this study is to explore the hybridization of NWP and observations using a streamlined neural network model. 
This ``hybrid'' neural network model that integrates NWP short-term forecasts with local ground station data is designed to process three distinct sets of MeteoNet input data: stations data, AROME predictions, and ARPEGE predictions. 
It is illustrated in Fig. \ref{fig:neural-network} and precisely described in the next section.
The model input feature variables $\bX^s_{d,t}$ can be therefore defined as 
$$\bX^s_{d,t} = \Big[\bFF^s_{d,t},\bGG^s_{d,t},\bHH^s_{d,t},\bC^s_{d,t},\bD^s \Big]$$
where, as explained below, $\bFF_{d,t}^{s}$,$\bGG_{d,t}^{s}$ and $\bHH_{d,t}^{s}$ are feature tensors built from respectively the ground station data, AROME and ARPEGE predictions at some given day $d$,  time $t$ and corresponding to some given site $s$. The tensors $\bC^s_{d,t}$ and $\bD^s$ bring deterministic information like the day of the year, the time of the day, the location of the station and its neighboring stations (see below).
Since NWP forecast data is only available for \( 0 \leq t \leq 24 \), the set of available features is constrained by the choice of the maximum forecast horizon, \( h_H = 6 \) hours. Consequently, for each day \( d \), the initial time \( t \) is restricted to the range \( t = 0,1, \dots, 17 \) spanning the hourly intervals from 0:00 AM to 5:00 PM.  It should also be noted that, prior to being input into the model, all tensors are standardized component-wise using their mean and variance computed from the training subset.

Each input component for a given location—that is, for any site referenced as a “ground station” in the MeteoNet dataset—is described more precisely below. These features are summarized in Table \ref{tab:input_data}.

\begin{table*}[h]
	\centering
	\caption{Summary of the MeteoNet input data used in the model $\cM$ or $\cM'$}
	\setlength{\tabcolsep}{3pt} 
	\small 
	\begin{tabular}{cccc}
		\hline\hline
		Input type & \multicolumn{1}{c}{Space and time grids}& \multicolumn{1}{c}{Weather fields}   &\multicolumn{1}{c}{Input shape}                                                                      \\ \hline
		Stations  & \begin{tabular}[c]{@{}c@{}}- Spatial grid: 11 locations (station + 10 neigh.) \\ - Time grid: Current+6 past hourly values  \end{tabular}                                                                                                                                                                   & \begin{tabular}[c]{@{}c@{}}- Wind u,v  (m/s) \\ - Temp. (K)  \end{tabular}    & $(7,33 )$                                                                                                                      \\ \hline
		AROME   & \begin{tabular}[c]{@{}c@{}}- Spatial grid : $(11 \times 11)$  \\ - Time grid : All 6 h ahead predictions \end{tabular}                                                                                                                                                                       & \begin{tabular}[c]{@{}c@{}}- 2 m Temp. (K) \\ - 2 m Rel. humidity (\%)\\ -  Wind u,v (m/s) \\ - Mean sea lev. Press. (Pa)\\ - \end{tabular} & $(6,5,11,11)$\\ \hline
		ARPEGE & \begin{tabular}[c]{@{}c@{}}- Spatial grid  : $(7 \times 5 \times 5 )$ \\ - Time grid : All 6 h ahead predictions \end{tabular} & \begin{tabular}[c]{@{}c@{}}- Temp. (K)\\ - Wind u,v (m/s) \\  - Pressure (Pa)\\ \end{tabular}  & $(6,4,7,5,5)$                   \\ \hline
	\end{tabular}
	\label{tab:input_data}
\end{table*}

First, $\bFF_{d,t}^s$, associated with weather ground station data, is composed of 3 different variables: the wind speed data ($u$,$v$ components) and surface temperature data. Other meteorological variables, such as atmospheric pressure and humidity, exhibit a significantly lower availability of data. Consequently, we have chosen to focus exclusively on wind speed horizontal components and temperature as primary features. For each site, we incorporated these three variables as measured at the ten nearest surrounding weather stations where data were available, as well as at the site itself. Accordingly, at any given point defined by $(s,d,t)$, the total number of station-related input variables amounts to $N_C = 33$. Furthermore, for each initial time $t$, we consider not only the observations at time $t$ but also those recorded at six preceding time steps, corresponding to $t-1$ through $t-6$, with each step representing a six-hour interval. It is important to note that the MeteoNet dataset provides ground station measurements at a temporal resolution of six minutes. However, our experiments have shown that increasing the time resolution does not lead to any significant improvement in forecasting performance within the considered model. Therefore, to maintain model simplicity and minimize the number of parameters, the analysis is restricted to data at an hourly scale.
The number of time periods (past or present) is then $N_H=7$, so that the overall $\bFF_{d,t}^{s}$ input tensor shape  is $(7 \times 33)$.

In order to build $\bGG_{d,t}^s$, among all AROME prediction data, we select wind eastward and northward components at 10 $m$, mean-sea-level pressure, 2~m temperature and relative humidity so that the number of input variables is $N_C = 5$. We keep only data around the considered ground station position, i.e., we consider the set of $11 \times 11$ AROME grid points centered around the ground station, namely if $i_0,j_0$ are the coordinates of the grid point closest to the target ground station $s$, then the considered AROME predictions are those within a spatial extent of $\pm 0.125 ^\circ$ latitude/longitude around this grid point.
Moreover, we keep all prediction horizons $t+1,t+2,\ldots, t+6$ so that $N_H = 6$. The  shape of the tensors $\bGG^{s}_{d,t}$ is therefore $(6 \times 5 \times 11 \times 11)$. 

The $\bHH^{s}_{d,t}$ tensor, which contains the data from the ARPEGE NWP model, is constructed along the same line: we keep only wind eastward and wind northward components together with temperature at all 7 isobar levels and we also consider the pressure levels at all 7 heights. As far as prediction horizons are concerned, we consider the same 6 next hours set of horizons as for AROME. When the horizon is not available,  we consider the closest horizon multiple of 3 hours (indeed, ARPEGE predictions are not available on a regular time grid, as the time step varies from 1 hour in the morning to 3 hours in the afternoon). Concerning the grid size, we use all grid points within an interval $\pm 0.2^\circ$ around the closest point to the station. Notice that this spatial extent is larger than former AROME one but, due to the lower resolution of ARPEGE, it corresponds to a smaller ($5 \times 5$) grid size.
Accordingly,
the ARPEGE input tensors have a shape $(6 \times 4 \times 7 \times 5 \times 5)$.

Finally the tensors $\bC^s_{d,t}$, contains a set of temporal and static features are also provided at station $s$, day $d$ and time $t$.  As in, e.g., \citet{BaileMuzy2023}, the initial time $t$ and the day index $d$ are encoded using $\cos$ and $\sin$ functions, namely time $t$ is represented by $(\cos(\frac{2 \pi t}{24}), \sin(\frac{2 \pi t}{24}))$ and $d$  by  $(\cos(\frac{2 \pi d}{365}), \sin(\frac{2 \pi d}{365}))$. Information about the location (latitude, longitude and altitude) of the station $s$ is also provided. An encoding of the position of each of the 10 neighboring sites of $s$, $\bD^s$ with respect to $s$ is also used.

\subsection{Model architecture}
\label{sec:model_arch}
\begin{figure*}[h!]
        \vspace*{0.2cm}
	\hspace*{0.5cm}
\includegraphics [width= 0.9\textwidth]{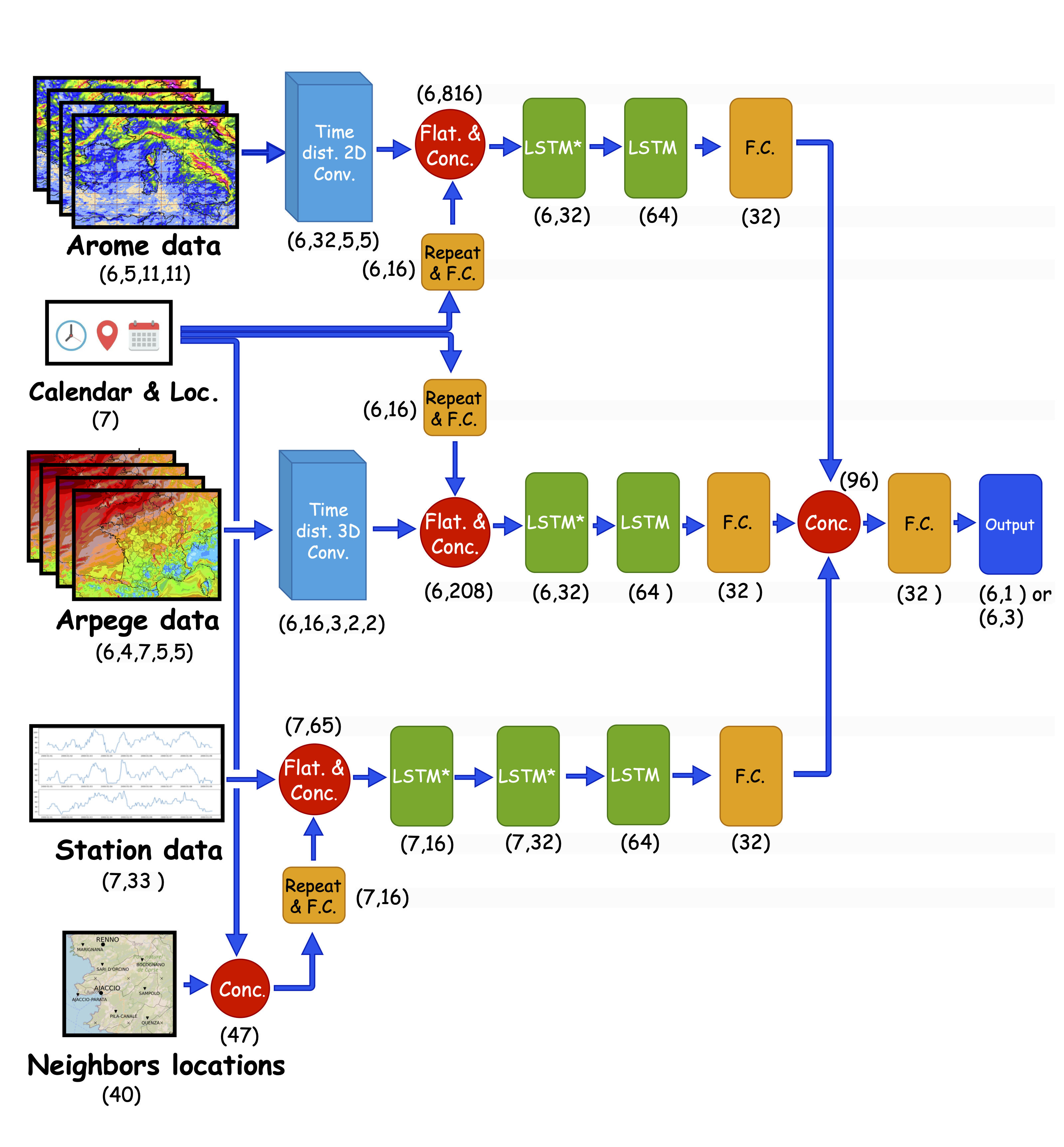}
	\vspace*{-0.2cm}
	\caption{Schematic representation of the neural network architecture used to process station, AROME, and ARPEGE data. The output shape of each layer is indicated below the corresponding block. "F.C." denotes a fully connected layer, "Flat." stands for a flattening operation, "Conc." indicates a concatenation of features, and $LSTM^*$ refers to an LSTM layer that returns the full output sequence rather than only the final hidden state.}
	\label{fig:neural-network}
    \vspace*{0.0cm}
\end{figure*}

The overall model architecture, illustrated in Fig.~\ref{fig:neural-network}, is composed of three branches devoted to handling the three different input features described previously. Rather than implementing a state-of-the-art architecture, this study adopts a simple neural network design that leverages conventional convolutional and recurrent layers for spatial and temporal feature extraction \citep{deeplearning2016}. While the core architecture remains intentionally straightforward, we also experiment, as detailed in  Section~\ref{sec:subsets}, with several architectural variants such as alternative temporal encoders and adjustments to layer configurations. Identifying more advanced architectures and optimizing model performance is left for future research.

One can see in Fig. \ref{fig:neural-network} that the branches that process AROME data ($\bGG^{s}_{d,t}$)  and ARPEGE data ($\bHH^{s}_{d,t}$) inputs begin with a time-distributed convolution (represented by blue 3D boxes), which applies the same spatial (2D for AROME and 3D for ARPEGE) convolution operation independently to each time step in a sequence of inputs, treating each time step as a separate sample. This allows the model to extract spatial features along the temporal axis while maintaining the temporal dimension for subsequent processing. For each of the $6$ times indices, the output of convolution layer is then flattened in order to provide a 2-dimensional tensor (red circular boxes on the sketch, of shape $(6,816)$ and ($6,208)$ for AROME and ARPEGE branches respectively), such that they have the same number of dimension as the Station branch (whose input shape is $(7, 33)$). 
Each of the static tensors $\bC^s_{d,t}$ and $\bD^s$ are processed by a fully connected layer which encodes this information within a 16-dimensional tensor. 
These tensors are then repeated at each of the input times (7 times for station data and 6 times for AROME and ARPEGE branches) and then concatenated to former input tensors or outputs of convolution layers. The so-obtained tensors are then processed through 2 or 3 stacked 
``Long Short-Term Memory'' (LSTM) layers, a recurrent neural network architecture that is widely used to process sequential information \citep{deeplearning2016} and which is represented by a green rectangle in Fig. \ref{fig:neural-network}. The outputs of all branches, as provided by  
a last fully connected layer, are then concatenated into a single tensor to be encoded by a final fully connected layer and then by the output layer. 

As described previously, for the deterministic version of the model, $\cM$ the output layer is, in agreement with Eq. \eqref{eq:mod_gen} also a fully connected layer with a rectified linear unit (ReLU) activation that returns a vector of 6 positive values representing the components of $\hbV^s_{d,t+\bH}$ the predicted wind speeds at horizons $h=1,2,\ldots,6$ hours. 
In the case of probabilistic prediction, as described by the model $\cM'$ in Eq. \eqref{eq:mod_gen_p}, 
the last layer is a fully connected layer which outputs a $(n,6)$ tensor, where $n=3$ is the number of parameters necessary to define the M-Rice probability distribution described in \ref{app:M-Rice}.

We will also consider the restriction of the model $\cM$ to a single type of input, i.e., one that considers specifically $\Big[\bFF^s_t,\bC^s_{d,t},\bD^s \Big]$, $\Big[ \bGG^s_t,\bC^s_{d,t} \Big]$ or $\Big[ \bHH^s_t,\bC^s_{d,t} \Big]$ and their corresponding branch. When $\cM$ is reduced to its ``ground station'', ``AROME'' or its ``ARPEGE'' branch, the corresponding models will be denoted as respectively {\bf $\cM_{GS}$} model, {\bf $\cM_{AR}$} model and {\bf $\cM_{AP}$} model.

 Regarding the number of parameters to be learned in each model, $\cM, \cM'$, as previously described, consists of 262k parameters, while $\cM_{AR}$, $\cM_{AP}$, and $\cM_{GS}$ contain 138k, 84k, and 39k parameters, respectively.

\subsection{Baseline models}
\label{sec:baselines}
In order to compare the performance of previous neural network models with alternative approaches, we consider various simple or natural baselines.
The first model we consider corresponds to the raw output directly provided by the NWP regional model, namely the AROME model. For a given site $s$, some given day $d$ and time $t$,  at the lead time $h$, we use the wind speed values from the closest grid point in AROME at time $t+h$. This will be referred to as the ``Raw AROME'' prediction. Although, as  discussed in the introduction, such raw predictions can often exhibit some bias and can be improved through different post-processing approaches, it is the simplest baseline model one can use for reference.
Along the same line we also consider ``Raw ARPEGE'' as the prediction provided by the global ARPEGE model
at the grid point closest to the considered site. Given that AROME predictions are downscaled from ARPEGE, it is expected that AROME will deliver superior performance over ARPEGE.
Another model that is usually considered as a reference model for surface wind speed is the so-called ``Persistence'', which consists in considering the last measured value as the best forecast. Mathematically, this amounts to assuming that the wind speed is a Martingale random process. Finally, for the sake of completeness,  we consider the class of linear models as last comparative models. Such models can notably be considered as the most basic neural network with a single fully connected layer with a linear output and which takes the past observed wind speed values as input.  This model will be referred to as ${\cal L}$ (for linear) model.
In section \ref{sec:Corsica_ft}, we also consider, for comparison purpose 
on specific sites in Corsica, the so-called $\cal N$-model which is a neural-network model proposed in \citet{BaileMuzy2023} which is based on ground observations at a given site and its nearby stations.

\subsection{Training, validation and test sets, fine-tuning, model's parameters and  hyperparameters selection.}
\label{sec:subsets}

As detailed in Sec.~\ref{sec:data}, the MeteoNet dataset spans a three-year period from 1 January 2016 to 31 December 2018. The input data for the models $\cM$ and $\cM'$ are constructed from (normalized) values associated with each station $s$, day $d$, and initial time $t$ (with $t \leq 17:00$ UTC, expressed in hours). We refer to each triplet $(s, d, t)$ as a \emph{key}. Stations included in this study are those with at least 2000 valid keys—i.e., at least 2000 triplets with no missing values among the features listed in Table~\ref{tab:input_data}. This filtering results in 278 stations (out of 478) being retained, yielding a total of approximately $N \simeq 3.5 \times 10^6$ valid keys. To ensure reliable model training and evaluation, this global sample is divided into three mutually exclusive subsets: 70\% for training, 20\% for validation (used to monitor training and tune hyperparameters), and 10\% for testing (used for final model evaluation). Importantly, to avoid potential temporal data leakage—i.e., the risk that highly correlated samples appear in both the training and validation/test sets—the partitioning is performed at the \emph{day} level. That is, all keys from a given day $d$ are assigned exclusively to one of the three subsets, ensuring that no overlapping or closely correlated data points are shared across splits. This design prevents situations where, for example, data from station $s$ at day $d$ and time $t$ is in the training set, while data from the same station or a nearby one at the same or adjacent times appears in the validation or test set. Moreover, as detailed in Sec.~\ref{sec:model_inputs}, we only consider initial times up to 5:00~PM (17:00 UTC) each day. This additional constraint ensures that input sequences used for prediction on day $d$ do not incorporate data from the following day $d+1$ and remain sufficiently distant from the previous day $d-1$, further mitigating the risk of information leakage across subsets.


In addition to global training, we also experiment with fine-tuning the model to optimize its performance for specific stations. Fine-tuning a neural network involves taking a pre-trained model and continuing its training on a given subset of data. This allows the model to adapt to a particular site, to capture local meteorological patterns and station-specific biases that may not be fully represented in the global training. This improves the predictive accuracy for a given domain or task \citep{quinn2019dive} which can be particularly beneficial in operational contexts. Fine-tuning is often used when the general structure of the model is suitable but improvements are needed for specific data characteristics. It is different from the so-called ``transfer learning'' that refers to leveraging a model pre-trained on a large dataset and reusing its learned features for a new task. Transfer learning typically involves replacing or adding new layers to adapt the model to the new problem, while fine-tuning focuses on further optimizing the pre-trained model for better performance on the target dataset. Unlike transfer learning—which typically involves modifying the model architecture to suit a new task—fine-tuning retains the original structure and focuses on refining the learned parameters for the target dataset. To fine-tune the model for a given station $s$, we initialize it with the parameters obtained from the global training phase. The model is then further trained using the same loss function defined in Eq.~\eqref{def:logS_multi}, but restricting the training data to samples associated with station $s$ only.  
To mitigate overfitting, the fine-tuning procedure maintains the same data partitioning as in the global training phase, but restricts the training, validation, and test sets to samples associated with the target station $s$.

Hyperparameters are settings that govern both the learning dynamics and architectural configuration of a neural network.  Unlike model parameters, which are optimized during training, hyperparameters must be set prior to training. In this study, these include the spatial resolution of input data (e.g., AROME and ARPEGE grids), the number of units in LSTM and fully connected layers, the number of recurrent layers, dropout rates, batch sizes, and the learning rate of the optimizer. 
In this study, hyperparameter selection followed a two-stage strategy. First, we performed an initial manual exploration to establish a stable and reasonably well-performing baseline configuration. This initial phase combined the use of training hyperparameter values commonly adopted in the literature (e.g., typical values for learning rate, batch size, and dropout rate) with a manual exploration of various architectural alternatives to assess their impact on model performance. Specifically, we tested:
\begin{itemize}
    \item[-] Doubling the resolution of AROME and ARPEGE input data.
    \item[-] Adjusting the number of units and layers for encoding static inputs (e.g., station position, time, and date).
    \item[-] Halving or doubling the number of units in fully connected layers within each branch and in the final layers after merging.
    \item[-] Modifying the number of LSTM layers or their hidden units.
    \item[-] Substituting LSTM layers with alternative temporal encoders, such as GRU, Temporal Convolutional Networks (TCNs), or Transformer encoders, in the relevant branches.
\end{itemize}
None of these higher-level alternatives led to a consistent improvement in performance over the architecture presented in Section~\ref{sec:model_inputs} (Fig.~\ref{fig:neural-network}), which we retained for all reported results.
In the second stage, we conducted a systematic grid search to fine-tune training hyperparameters around the baseline configuration. The grid search explored four key parameters: learning rate in $\{0.001, 0.01\}$, batch size in $\{256, 512, 1024\}$, dropout rates (applied to all LSTM layers) in $\{0.01, 0.02, 0.2\}$  and the number of LSTM units in the largest LSTM layers  $\{16,32,64\}$, with neighboring layers using half or a quarter of these values as indicated in Figure \ref{fig:neural-network}. Each model was trained for a fixed number of steps (specifically, $10$ epochs). The hyperparameter values selected through this process are reported in Table~\ref{tab:learning_parameters}.
It is noteworthy that the limited hyperparameter search space used to identify the best-performing configuration reduces the risk of overfitting to the validation set, thereby suggesting minimal differences between validation and test errors. This expectation is supported by the close agreement observed between validation and test performance, which differ by only a few percentage points—indicating robust generalization of the selected hyperparameters.

\begin{table}[t]
	\caption{Some of learning Hyperparameters used in this study}
	\label{tab:learning_parameters}
	\begin{center}
		\begin{tabular}{ccrrcrc}
			\hline\hline
			Parameter &  Value\\
			\hline
			Optimizer & Adam  \\
			Learning rate & 0.001 \\
			Batch size & 1024 \\
			LSTM Dropout level & 0.02 \\
			\hline
		\end{tabular}
	\end{center}
\end{table}

All models presented in this study, including both deterministic and probabilistic architectures, were implemented and trained using the Keras API with TensorFlow \citep{tensorflow2,keras}. The TensorFlow framework provides efficient support for GPU acceleration and large-scale data pipelines, which was essential for training the models on the MeteoNet dataset and for rapid evaluation during fine-tuning and validation phases.
All training were performed using a single Nvidia V100 GPU. The total training time for models $\mathcal{M}$ and $\mathcal{M'}$, that is, the most computationally demanding among models considered in this study, is approximately 3 hours (about 220 seconds per epoch with $\approx 50$ epochs needed for convergence). As typical for NNs, training is performed only once. Afterward, the model can generate predictions on a new batch of data in just a few seconds ($\approx\,5 \,s$ including the time needed for model loading and set-up).

\section{Results}
\label{sec:Results}
%
%
%
%
%
%
\subsection{Hybrid model vs various baselines over all station sites}
\label{sec:SE}

\subsubsection{Comparison with basic models: raw AROME prediction and persistence models}
\label{sec:comp1}
\begin{table}[h]
	\caption{RMSE relative improvement of $\cM$ with respect to baseline models at horizons $1$ and $6$ hours.}\label{tab:rel_perfs}
	\begin{center}
		\begin{tabular}{ccccrrcrc}
			\hline\hline
			& Baseline  / horizon  & 1h & 6h\\
			\hline
			
			& Persistence & 6.2 \% & 45 \% \\
			& Linear & 2 \% &  37 \%  \\
			& Raw AROME & 37 \% & 28 \% \\
			& Raw ARPEGE & 49 \% & 41 \% \\
			
			\hline
		\end{tabular}
	\end{center}
\end{table}

\begin{figure}[h]
	\centering
	\includegraphics[width=1.0\linewidth]{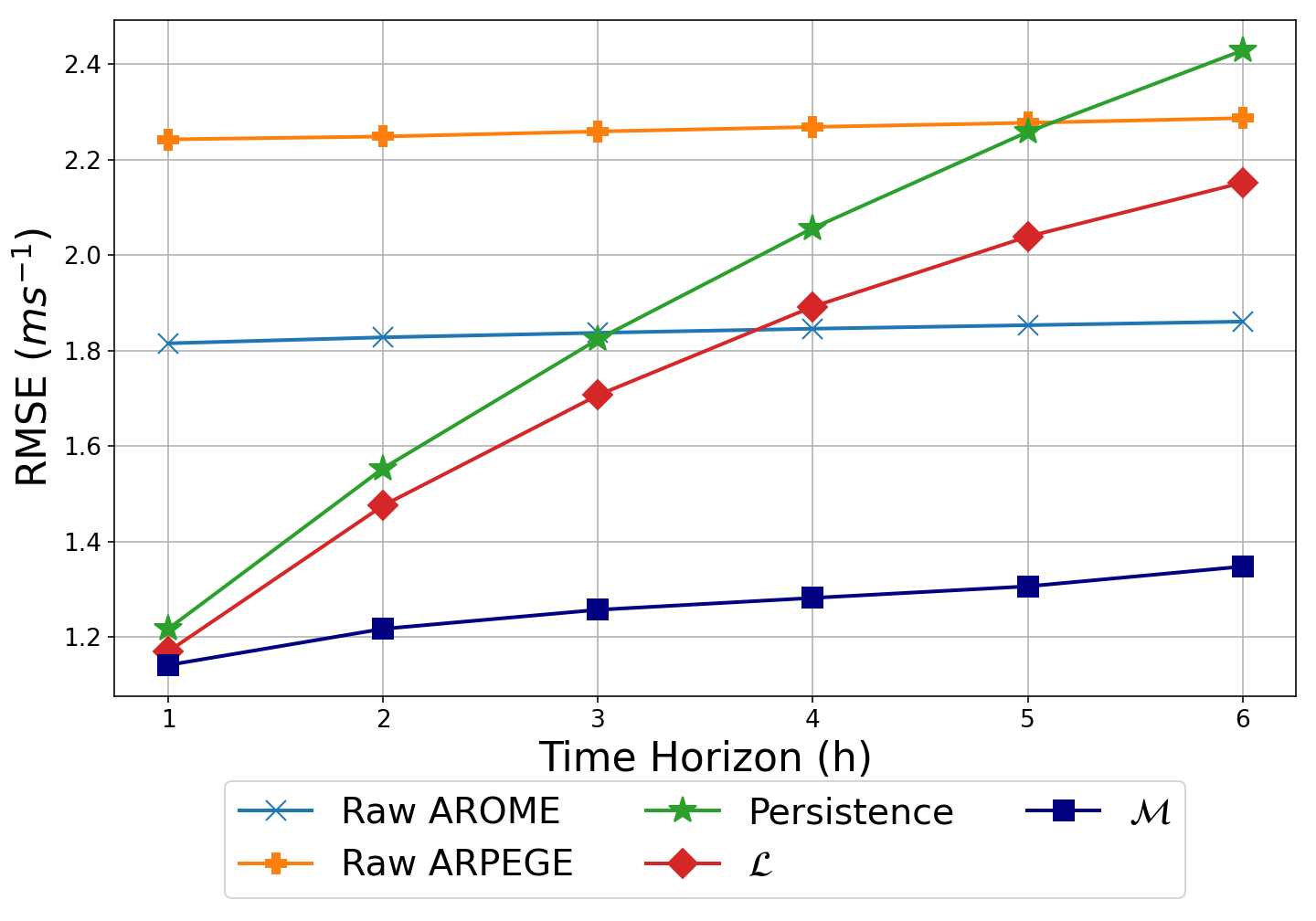}
	\caption{Comparison of the performance of the $\cM$ model (symbols (\textcolor{blue!45!black}{$\blacksquare$})) against various baseline approaches described in Secs. \ref{sec:model_arch} and \ref{sec:baselines}. The Root Mean Square Error (RMSE), expressed in $ms^{-1}$ and defined in Eq.~\eqref{eq:rmse}, is computed over the test dataset and reported for each forecast horizon ranging from 1 to 6 hours. 
	}
	\label{fig:Baseline_RMSE}
    \vspace*{-0.2cm}
\end{figure}

The comparative performances in terms of RMSE of the $\cM$ model and the various baseline models are reported in Fig. \ref{fig:Baseline_RMSE}.
These results correspond to the RMSE evaluated over the test set for all 278 stations in the MeteoNet dataset for prediction horizons $h = [1,2,3,4,5,6]$ hours. One can see that the $\cM$ model significantly outperforms all baselines for all horizons,
as highlighted in Table~\ref{tab:rel_perfs} for horizons $1$ and $6$ hours. We see that
Linear ($\cal L$) and Persistence models based on station data provide good performances at very short time horizons
but quickly become less reliable as the forecasting horizon increases. On the other hand, as anticipated in the discussion of Sec. \ref{sec:prediction}, we observe that raw NWP predictions (AROME and ARPEGE) errors hardly vary with horizon.  We also observe that, as expected, the regional, high-resolution AROME model performs better than the global ARPEGE model. AROME becomes better than a regression of station data at horizons greater that 3 $h$. However, both models are surpassed by $\cM$, even at largest time horizons. This clearly demonstrates how $\cM$ can greatly improve the prediction accuracy of raw NWP forecasts by leveraging multiple sources of input data and correcting bias inherent to these models. 

\begin{figure*}[h]
	\centering
	\includegraphics[width=1.0\linewidth]{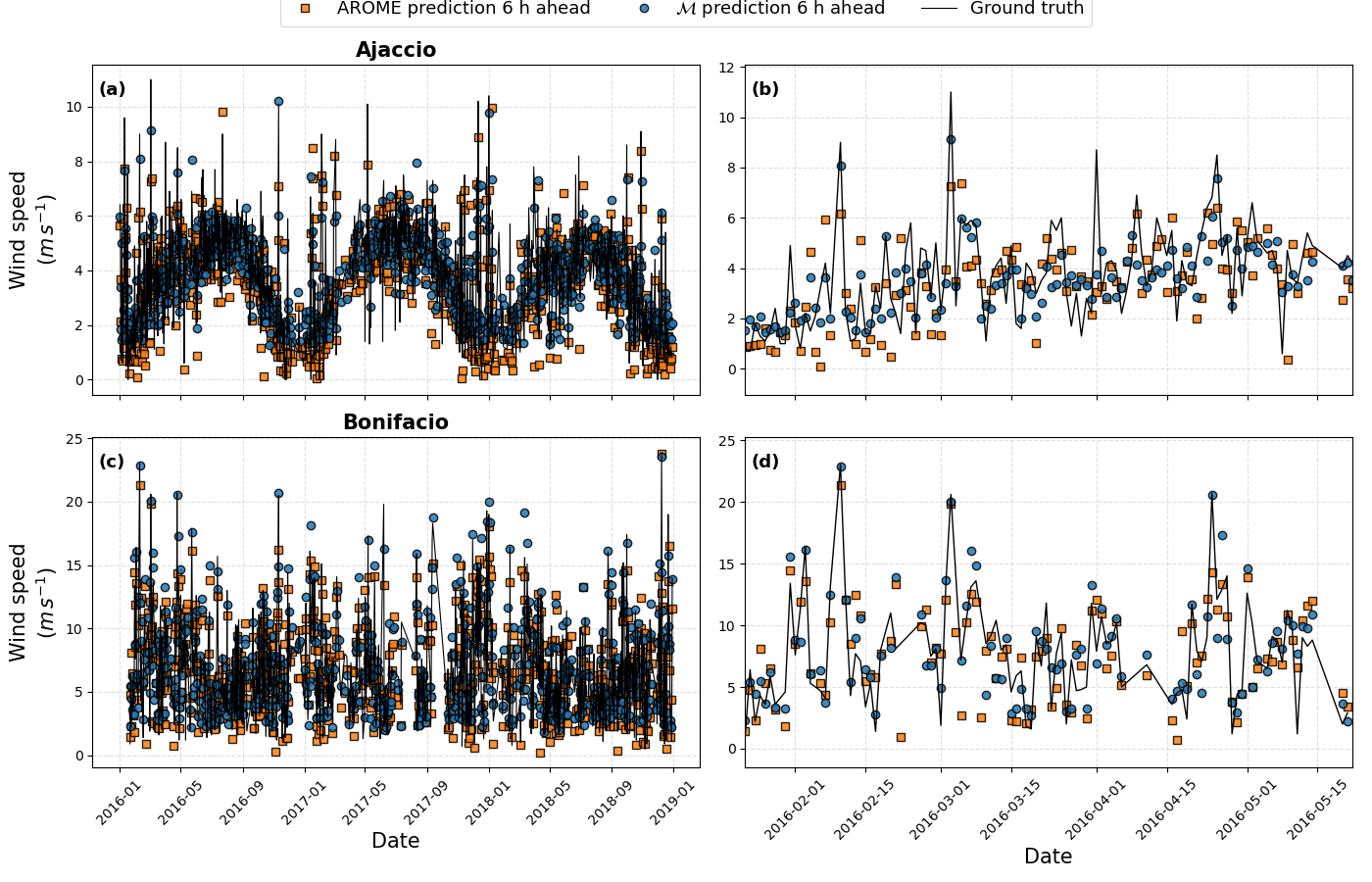} 
	\caption{Illustration of the $\cM$ (symbols  (\textcolor{blue}{$\bullet$})) model and the raw AROME forecasting performances (\textcolor{orange}{$\blacksquare$}) in Ajaccio and Bonifacio for horizon $h$ $=$ $6$ hours ahead prediction of wind speed value at valid time $t_h$ $=$  $t$ $+$ $h$ $=$  $1100$ UTC each day. Measured wind speed value as indicated by the black solid lines. Panels (a) and (c) display daily forecasts over the full 2016–2018 period, while panels (b) and (d) provide a zoomed view of a representative 4-month sub-period (15 January 2016 to 15 May 2016) to enhance visual clarity.  Although the overall performance of the two models appears comparable in (a) and (c), the raw AROME forecasts exhibit a slight bias toward lower wind speeds. One can also observe strong seasonal effects in Ajaccio where breeze regimes are much more important than in Bonifacio. The zoomed views in (b) and (d) confirm that the $\cM$ model outperforms the raw AROME forecasts quite frequently, on a day-to-day basis, indicating consistent statistical superiority.}
	\label{fig:Example_RMSE}
    \vspace*{-0.2cm}
\end{figure*}

For illustrative purposes, we plot in  Fig. \ref{fig:Example_RMSE} the observed wind speed together with model predictions at two different ground station sites in Corsica: Ajaccio and Bonifacio. In both subplots, we present the values of 6-hour ahead forecasts from the AROME and $\cM$ models at a fixed valid time of day $t_v = $ 1100 UTC (the corresponding initial time is therefore $t = t_v - h$ = 0500 UTC) for all days over the 2016–2018 period. For the sake of clarity, the plots include all three sub-periods (training, validation, and test subsets), which may result in slightly overestimated model performance compared to its actual values. In Ajaccio, we observe a pronounced annual oscillation related to the coastal breeze regime, whereas in Bonifacio, wind speeds are more intermittent and wind peaks more intense. 
We observe that both models produce predictions that closely match the ground truth across all locations and time periods. Notably, model $\cM$ consistently outperforms AROME, a difference that becomes particularly evident in Figs. \ref{fig:Example_RMSE}(b)–(d), which provide a zoomed-in view over few months for clarity. One can see that raw AROME tends to slightly underestimate wind speeds, especially during the winter season at both sites (in Fig. \ref{fig:Example_RMSE}, orange squares are more frequently located below the black solid lines than the blue circles).
This is confirmed by statistics of the model performances (over the validation and test sets) at a 6-hour forecast horizon: the RMSE of model $\cM$ over Ajaccio and Bonifacio are respectively 1.13~m$\cdot$s$^{-1}$ and 2.07~m$\cdot$s$^{-1}$, while for raw AROME the RMSE values are 1.43~m$\cdot$s$^{-1}$ and 2.23~m$\cdot$s$^{-1}$. 
These improvements illustrate the need for systematic bias correction in raw numerical weather prediction (NWP) outputs, emphasizing the role of post-processing or hybrid modeling to enhance forecast accuracy.

\subsubsection{Model performance as a function of site mean wind speed and across valid times}
\label{sec:perfs_vs_ws}

This section investigates how the forecasting accuracy of model $\cM$ varies with local wind conditions and time of day, highlighting the influence of mean wind speed and diurnal cycles on prediction errors.

\begin{figure}[h]
	\centering
	\includegraphics[width=1.0\linewidth]{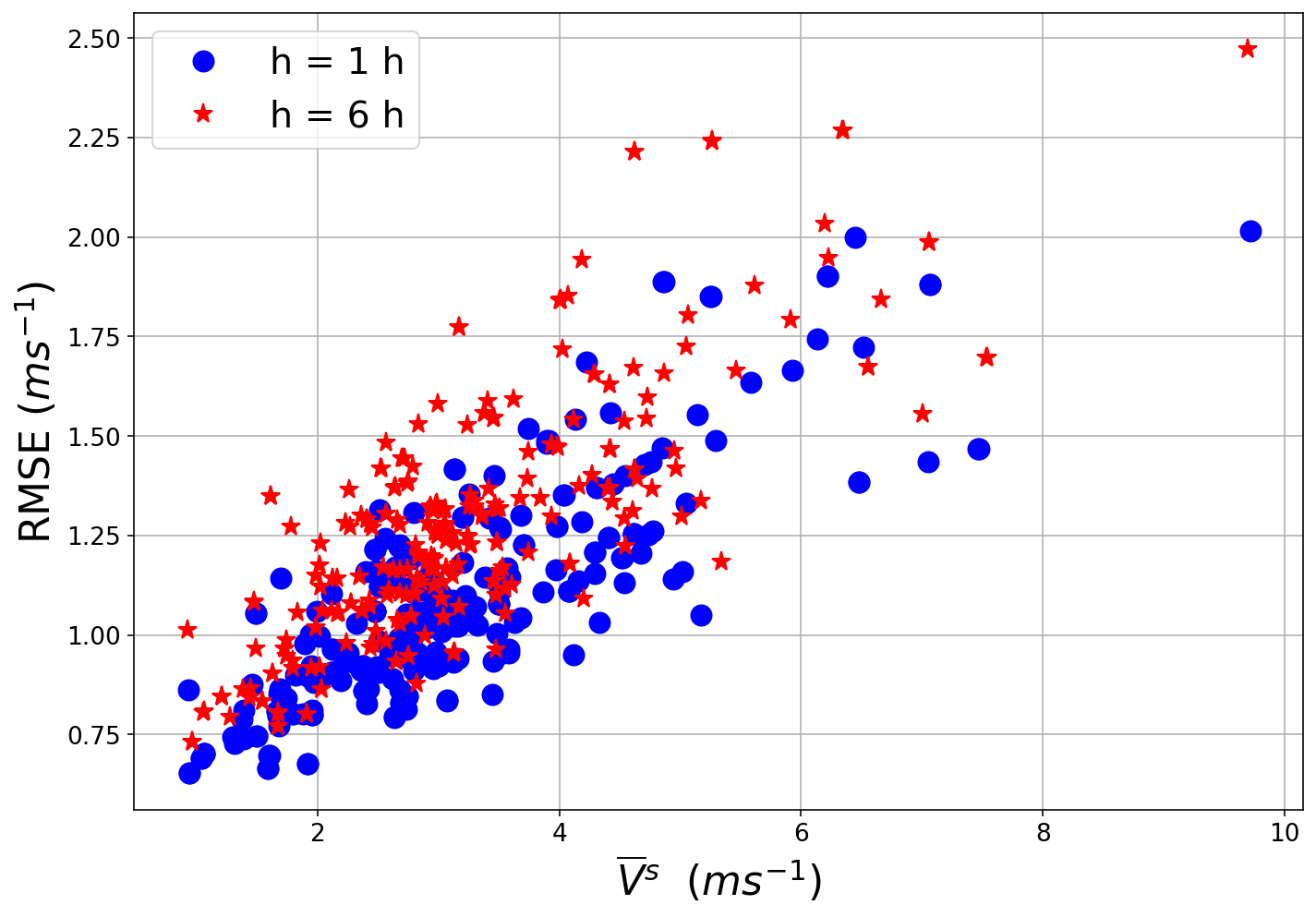}
	\caption{RMSE (in $m s^{-1}$) associated with the $\cM$ prediction as a function of the station mean wind speed value (estimated over 3 years) $\overline{V}^s$ for all sites for horizons $h$ = $1$ hour (symbols (\textcolor{blue}{ $\bullet$})) and $h$ = $6$ hours (symbols (\textcolor{red}{$\bigstar$})).}
	\label{fig:RMSE_vs_V}
    \vspace*{-0.2cm}
\end{figure}

It is first important to note that, beyond the average results presented in Fig. \ref{fig:Baseline_RMSE}, the RMSE-based forecasting performance depends of the $\cM$ model varies significantly across stations. This variability stems from the strong spatial variability in site-specific weather conditions, particularly mean wind speed. As shown in Fig. \ref{fig:RMSE_vs_V}, there is a clear linear dependence between RMSE and mean wind speed estimated at each site over the full 3 years period, with higher mean value leading to larger prediction error. 

\begin{figure}[h!]
	\centering
	\includegraphics[width=1.0\linewidth]{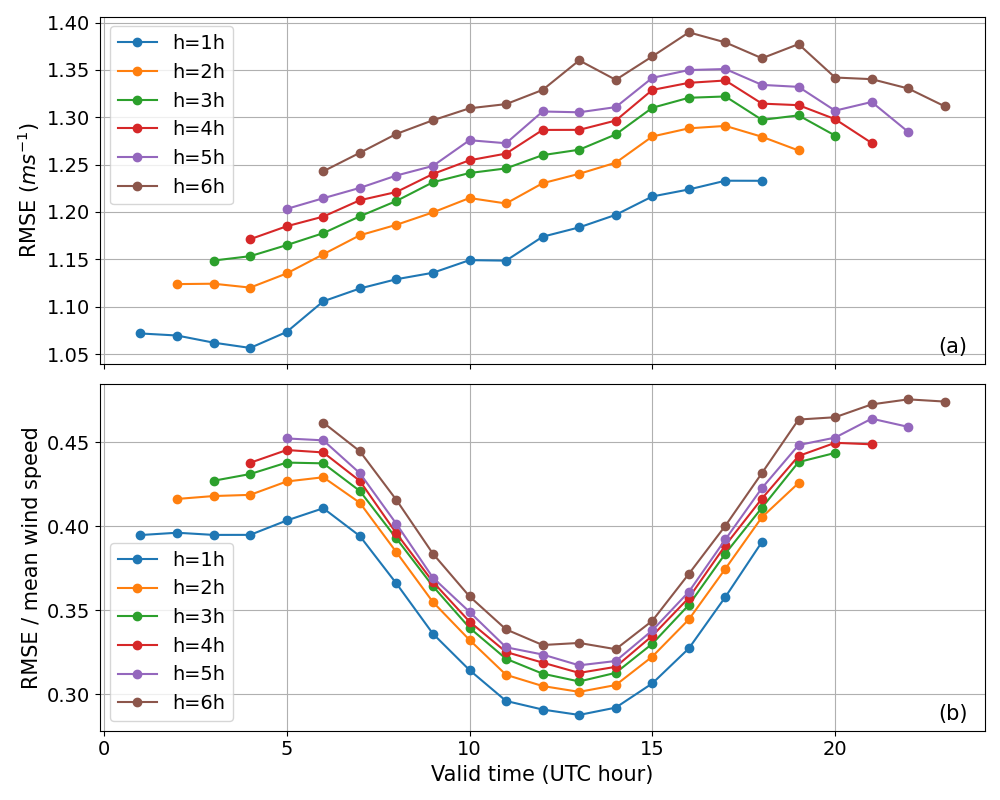}
	\caption{Diurnal variation of (a) RMSE and (b) relative RMSE (RRMSE) as a function of the valid time for model $\cM$ predictions at horizons $h$ ranging from 1 hour to 6 hours. Results are averaged over the 278 stations of the MeteoNet dataset. The RMSE (panel~a) slowly increases with valid time whereas the RRMSE (panel~b), normalized by the mean wind speed, shows a U-shaped pattern.}
	\label{fig:rmse_rrmse_diurnal}
    \vspace*{-0.2cm}
\end{figure}

Another key factor that likely influences the model’s forecasting performance is the \emph{valid time}---the targeted time of day for which the prediction is issued, as defined in Sec.~\ref{sec:prediction}. Figure~\ref{fig:rmse_rrmse_diurnal} illustrates how prediction errors, expressed as RMSE and Relative RMSE, vary with valid time. Here, valid time is the sum of the initial time and the forecast horizon, with results shown for horizons from $h = 1$~h to $h = 6$~h and averaged over the 278 ground stations in the dataset.  
In panel~(a), RMSE generally increases with forecast horizon: for a given valid time, earlier-issued forecasts tend to be less accurate. RMSE also shows a slight increase with valid time up to around 16:00--17:00~UTC, after which it gradually decreases. This  peak may reflect the fact that wind regimes typically become more turbulent and intermittent from early morning to mid-afternoon, reducing predictability. 
Nonetheless 
Panel~(b) presents Relative RMSE (RRMSE), computed by normalizing the RMSE at each valid time by the corresponding mean wind speed. This normalization highlights a pronounced U-shaped pattern: forecast skill as measured relatively to the mean velocity is lowest during the night and early morning, peaks around 14:00~UTC and then declines again in the late afternoon. This behavior primarily reflects the regional diurnal wind cycle, which follows a bell-shaped profile: winds are generally weak at night and in the early morning, then intensify through the afternoon under the influence of thermally driven sea-breeze circulations~\citep{diurnWind}, sometimes reinforced by synoptic flows such as the Mistral or Tramontane in the French Mediterranean.


\subsubsection{Comparison with single branch models}

\begin{figure}[h!]
	\centering
	
	\includegraphics[width=1.0\linewidth]{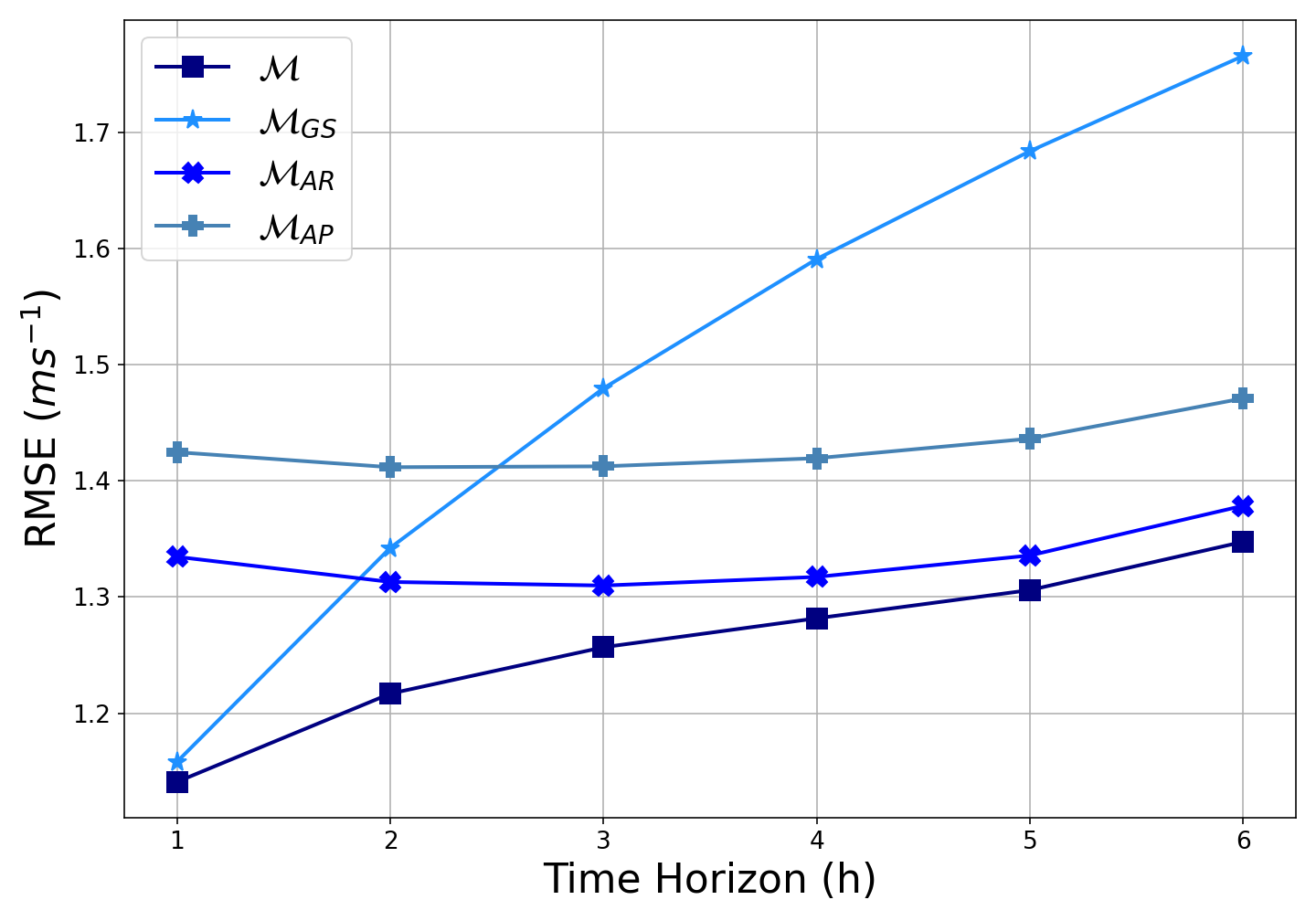}
	\caption{
    Comparison of the performance of the $\cM$ model  (symbols (\textcolor{blue!45!black}{$\blacksquare$}))  as respect to the $\cM_{AR}$  (symbols (\textcolor{blue}{$\protect \fatx$})), $\cM_{AP}$  (symbols (\textcolor{blue!45!green}{$\protect \fatplus$})) and $\cM_{GS}$ models  (symbols (\textcolor{cyan!80!blue}{$\bigstar$})). 
    The RMSE expressed in $ms^{-1}$ is computed over the test dataset and reported for each forecast horizon ranging from 1 to 6 hours.
	}
	\label{fig:M_RMSE}
    \vspace*{-0.2cm}
\end{figure}

In Fig. \ref{fig:M_RMSE} are displayed, as a function of the horizon, the RMSE performances of the $\cM$ model comparatively to $\cM_{AR}$, $\cM_{AP}$ and $\cM_{GS}$ models that respectively process
AROME, ARPEGE and Ground Station input data without other branches.
We first see that $\cM$ predictions are the best among all alternatives for all times horizons.
Comparing the results of Figs. \ref{fig:Baseline_RMSE} and \ref{fig:M_RMSE} shows that post-processing NWP inputs, whether AROME or ARPEGE, with a neural network model significantly reduces their error. Specifically, the RMSE for $\cM_{AR}$ predictions is reduced by approximately 27 \% compared to raw AROME predictions, while the RMSE for $\cM_{AP}$ predictions is reduced by about 36 \% compared to raw ARPEGE predictions.
The $\cM_{AR}$ model associated with AROME remains better than $\cM_{AP}$ associated with ARPEGE, even if the difference between the two models is reduced as compared to the raw predictions. In both models, as before and as expected from the discussion in Sec. \ref{sec:prediction}, the error only slightly depends on the time horizon. On the other hand, as remarked previously, the performance of $\cM_{GS}$, the model relying exclusively on past observations at ground station and its surrounding sites, strongly depends on the horizon and clearly decreases as the horizon increases.

\subsubsection{Feature importance analysis}
In the previous paragraph,  we have seen that the model $\cM$, which integrates all three data sources, consistently outperforms the best individual model at every forecast horizon. The model leverages the strengths of each model at each time horizon, optimizing performance across all forecasts. In order to assess the contribution of each feature to the reduction in final RMSE, we adopt the approach outlined in \citet{rasp2018}, which is based on a method originally proposed by \citet{breiman2001} for random forest models. This method simply consists, for a given input feature or group of features, in randomly shuffling the corresponding data among a given mini-batch sample. The main advantage of this method is its relative simplicity and the fact that it does not require retraining the model. This is not the case of the approach used by \citet{Marcille24}, which involves analyzing the impact of a specific subset of  features by removing it from the original input and run a new training phase within this configuration. 

\begin{figure}[h!]
	\centering
	\includegraphics[width=1.0\linewidth]{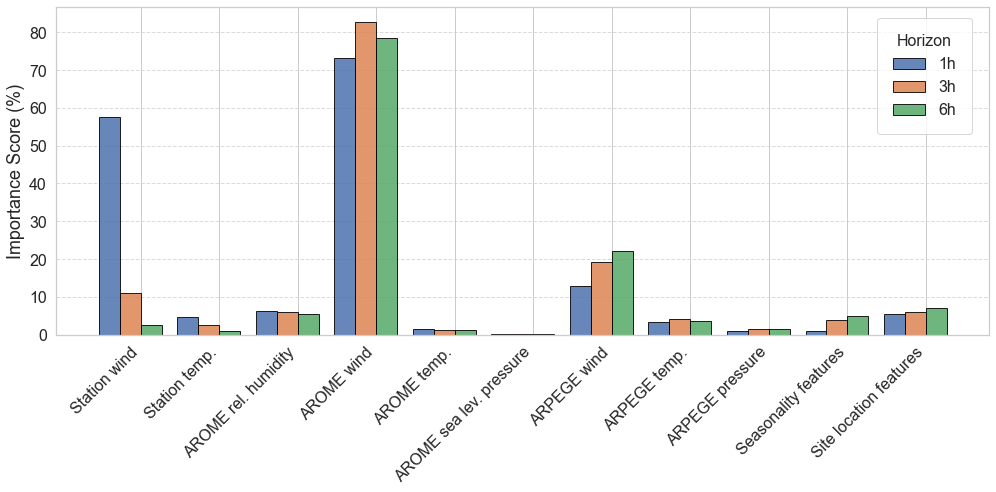}
	\caption{Importance score $S_F$ for each (class of) input feature $F$ at horizons 1 hour, 3 hours and 6 hours}
	\label{fig:fig_importance_score}

    \vspace*{-0.2cm}
\end{figure}

If ${\cal E}_{ref}[h]$ is the mean-squared error associated with the  reference model at a given horizon $h$ and ${\cal E}_{F}[h]$ is the mean-squared error at horizon $h$ obtained when randomly shuffling the input feature (or group of input features) $F$, then $S_F[h]$, an ``importance score'' of $F$ at horizon $h$ can be defined as:
$$
S_F[h] = \frac{\sqrt{{\cal E}_{ref}[h]}-\sqrt{{\cal E}_{F}[h]}}{\sqrt{{\cal E}_{ref}[h]}}  \; .
$$
In Fig. \ref{fig:fig_importance_score} are reported the importance score $S_F$ at horizons $h=1$, $h=3$ and $h=6$ hours for each feature in the ${\mathcal M}$ model computed on the test sample for the whole dataset. One can see that NWP and station wind components inputs are, by far, the most important features. Among other features, the relative humidity predicted by the AROME model and the temporal features (date, hour of the day) 
have the most significant impact of the prediction performances.
As far as horizons are concerned, there is no clear trend in AROME feature importance whereas ARPEGE wind feature become more important as the horizon increases. This is not surprising since ARPEGE is a global model that is designed to capture 
large scale features several days ahead. 
On the other hand, past station features become less important for 
longer horizons as compared to shorter ones.
Models based on observed past weather conditions at the target site and its surrounding ground stations are more likely to be useful at small horizons than for longer ones, as highlighted by Fig.~\ref{fig:M_RMSE}. 
It is noteworthy that for both AROME and ARPEGE temperature data have more impact on the prediction quality than pressure data. This may be reminiscent of the correlation of the wind components with thermal breezes as observed by \citet{Marcille24} for the meridional wind at the coastal station they analyzed.

Beyond its methodological interest, the feature importance analysis provides valuable insights for stakeholders and system operators. By quantifying the impact of each input feature on forecast accuracy, the importance scores help identify which data sources are most critical for reliable predictions. For example, the dominant role of NWP wind components and station observations confirms the operational relevance of maintaining high-quality measurements and forecasts from these sources. Similarly, the increasing importance of ARPEGE features at longer horizons suggests that global-scale forecasts are particularly useful for planning decisions further ahead in time. This information can guide decisions on sensor deployment, data validation, and resource allocation. It also supports transparency and trust in the model by offering a clear, quantifiable link between input data and output performance.  Moreover, such analysis can be leveraged to design more compact and efficient models by identifying and potentially excluding less informative inputs. This can reduce the number of model parameters, simplify training, and improve generalization. Simpler models are also more interpretable and computationally efficient, making them better suited for real-time forecasting applications and easier to maintain.

\subsubsection{The benefits of a probabilistic approach}

\begin{figure}[h!]
	\centering
	\includegraphics[width=1.0\linewidth]{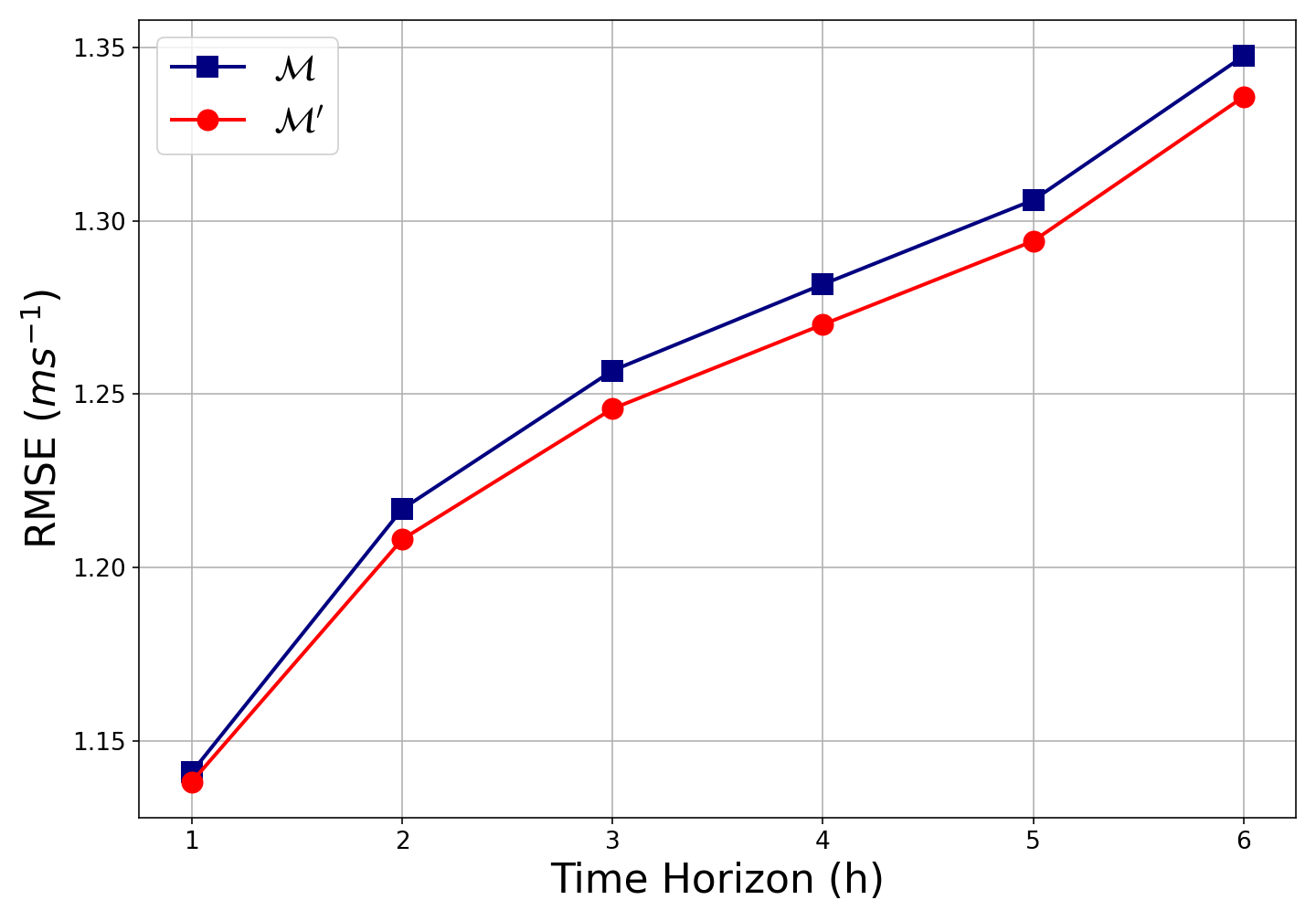}
	\caption{Comparison of model $\cM$ (symbols (\textcolor{blue!45!black}{\footnotesize $\blacksquare$})) and model $\cM'$ (symbols (\textcolor{red}{$\bullet$})) performance in their ability to provide a single point prediction. RMSE, in $m s^{-1}$, is plotted as a function of the forecasting horizon in hours.}
	\label{fig:DetPro_RMSE}
    \vspace*{-0.2cm}
\end{figure}

In Fig. \ref{fig:DetPro_RMSE}, we compare the RMSE observed on the test set for the $\cM$ model (blue ({\footnotesize $\blacksquare$}) symbols) and the $\cM'$ model (red ($\bullet$) symbols). Recall that the latter is designed to predict the parameters of the M-Rice probability distribution for the wind speed over the next six hours. At each time step, one can then compute, for any given forecasting horizon, the conditional mean using Eq. \eqref{eq:mean_mrice} and then the resulting MSE over the validation or the test samples. As it can be observed in Fig. \ref{fig:DetPro_RMSE}, whatever the considered horizon the probabilistic model ($\cM'$) provides slightly better forecasts than the deterministic version ($\cM$). To ensure the statistical significance of these results, we applied the Wilcoxon signed-rank test \citep{wilcoxon1992individual}, a non-parametric test for paired data that evaluates whether the median difference between two related samples, in this case, the forecast errors from $\cM$ and $\cM'$, is significantly different from zero. The test confirms that  $\cM'$ is more accurate, as the obtained  p-values are very close to zero ($p<10^{-20}$ for all horizons).

A priori, the model designed to optimize directly the empirical MSE ($\cM$) is expected to provide better results in terms of MSE as compared to any other one. Fig. \ref{fig:DetPro_RMSE} indicates that parametrizing directly the conditional mean is less efficient that trying to estimate it through the probability law it is associated with. This is even more true as the forecast horizon increases.
Moreover, as advocated previously, probabilistic forecasts offer valuable insights beyond a single-point prediction. They provide the entire probability distribution, offering the possibility to compute various risk measures and facilitating decision-making tailored to specific objectives. Even for single-point prediction, as considered in this paper, if one uses different metrics, the probability law can provide more efficient prediction than the MSE. For example, if one aims at minimizing another score instead of the MSE (e.g. the mean absolute error), there is no need to perform another training of the model since it suffices to compute, for each station and for each time $t$, the value that optimizes this score as the best forecast instead of the conditional mean value (e.g., the conditional median). Section \ref{sec:ext_winds}  will show that in the case of extreme wind occurrence prediction, high quantile values threshold exceedances provide a very good predictor which outperforms significantly the deterministic predictions.

\subsection{Focus on sites in Corsica}
\label{sec:Corsica}
In this section, our aim is to focus more specifically on a particular subset of locations, that is the 21 ground stations in Corsica that have sufficient available data for analysis. The precise locations and names of these stations are shown in Fig.~\ref{fig:sta_loc}(b). 

\subsubsection{Fine tuning the global model for each station}
\label{sec:Corsica_ft}
\begin{figure}[h!]
	\centering
	\includegraphics[width=1.0\linewidth]{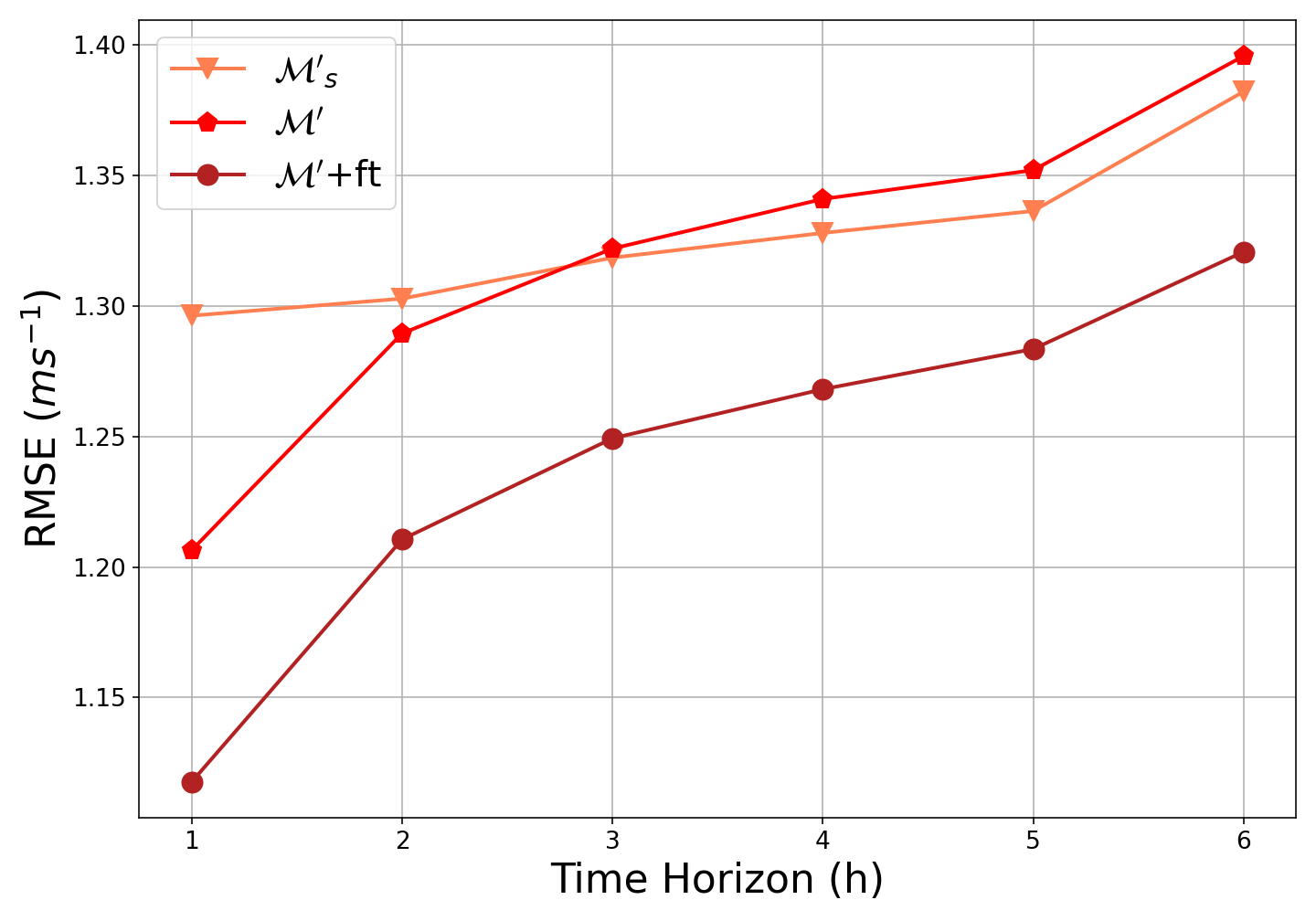}
	\caption{Comparison of the average RMSE (in $m s^{-1}$) over the 21 Corsican stations with and without fine-tuning. Symbols (\textcolor{red}{$\pentagofill$}) represent the $\cM'$ model as optimized globally (without any fine-tuning), (\textcolor{orange}{$\blacktriangledown$}) the average RMSE of the $\cM'_{s}$ model which is the $\cM'$ model fully optimized for each station without any peculiar initialization and the symbols (\textcolor{red!70!black}{$\bullet$}) represent the average RMSE of the fine-tuned $\cM'$ model.}
	\label{fig:fig_fine_tuning}
    \vspace*{-0.2cm}
\end{figure}
As noted in the Introduction, Corsica is a Mediterranean island characterized by a highly heterogeneous, complex, and mountainous terrain, subject to diverse weather conditions.
These factors contribute to significant spatial variability in wind regimes. Consequently, the performance of the model (e.g., $\cM'$), which has been globally trained using all stations, is expected to vary significantly across sites.
To address this challenge and enhance model performance, we explore the possibility of fine-tuning its parameters for each station according to the methodology described in Sec. \ref{sec:subsets}.
For the probabilistic model $\cM'$, fine-tuning is performed using only the data corresponding to the target station. The globally trained model is loaded to provide initial weights for the training phase, and the model is then trained using the same architecture and negative log-likelihood loss (Eq. \eqref{def:logS_multi}). This operation is computationally inexpensive and can be completed on a standard laptop within minutes.

Figure \ref{fig:fig_fine_tuning} presents, using symbols ($\bullet$), the RMSE for each forecast horizon, obtained as the square root of the averaged MSE of the fine-tuned model across the 21 stations reported in Fig. \ref{fig:sta_loc}(b). For comparison, the figure also includes the analogous RMSE obtained without model retraining, i.e., corresponding to the globally optimized model $\cM'$ (symbols ($\pentagofill$)). Is also shows the average RMSE of $\cM'_{s}$, the ``locally trained'' models that corresponds to the model $\cM'$ which is fully trained  specifically for each station without providing any peculiar initial weights (symbols ($\blacktriangledown$)).

The results clearly show that fine-tuning improves performance as respect to both the globally trained model and locally trained models. 
It is worth mentioning that the improvement of the fine-tuned model as respect to the global one is consistently observed for all stations and all horizons. The results are statistically significant as indicated by the Wilcoxon test, which for all horizons results in p-values of the order of $10^{-5}$ or smaller. 
The performance gain compared to the model $\cM'_{s}$, fully trained on data from a single station (symbols ($\blacktriangledown$)), is even more pronounced at shorter time horizons. Training $\cM'$ for a given station $s$ in isolation, without leveraging the optimal weights obtained from the full set of stations, leads to poorer performance on the validation and test subsets. This degradation can be primarily explained by overfitting, as the amount of training data available for a single station is significantly smaller than that of the entire Southeast station network.
\begin{figure}[h!]
	\centering
	\includegraphics[width=0.95\linewidth]{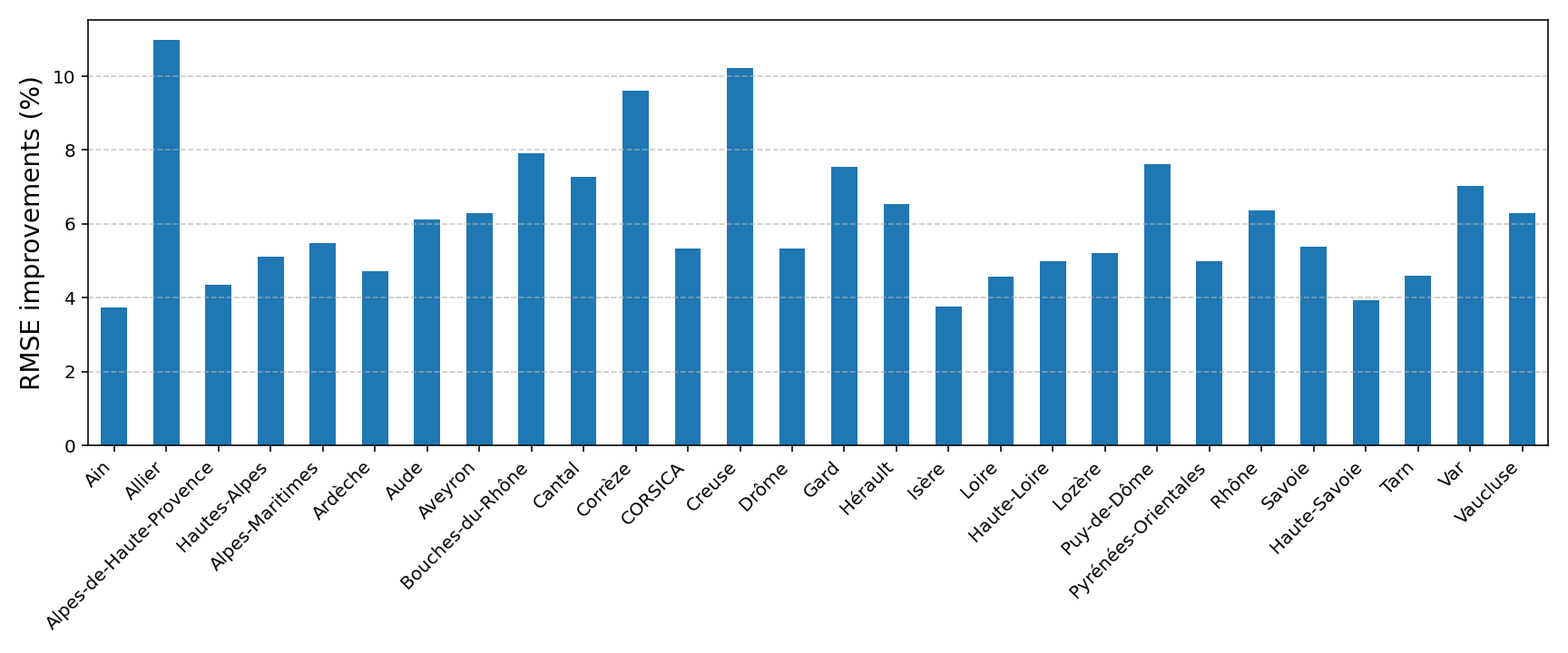}
	\caption{Relative RMSE improvement over the French administrative \emph{départements} included in the MeteoNet Southeast dataset. Results are averaged over all stations in each \emph{département}) and over all horizons. }
	\label{fig:RMSEimpSE}
\vspace*{-0.2cm}
\end{figure}

As a remark, the relevance of the fine-tuning procedure is not limited to the region of Corsica; in fact it can be applied to any subset of data from the original dataset. To demonstrate the general benefit of this approach, fine-tuning was tested over all French administrative \emph{départements}  included in the Southeast MeteoNet dataset, retaining only those stations for which data were of sufficient quality. The relative RMSE improvement, averaged over all the stations in the \emph{département} and over all forecast horizons, is illustrated in Figure \ref{fig:RMSEimpSE}. Observed improvements range between approximately $4\%$ and $11\%$.    

\subsubsection{Seasonal Performance Analysis and Comparison to model $\cal N$}
\label{sec:seas_perf}
\begin{figure}[h!]
    \centering
    \includegraphics[width=0.45\textwidth]{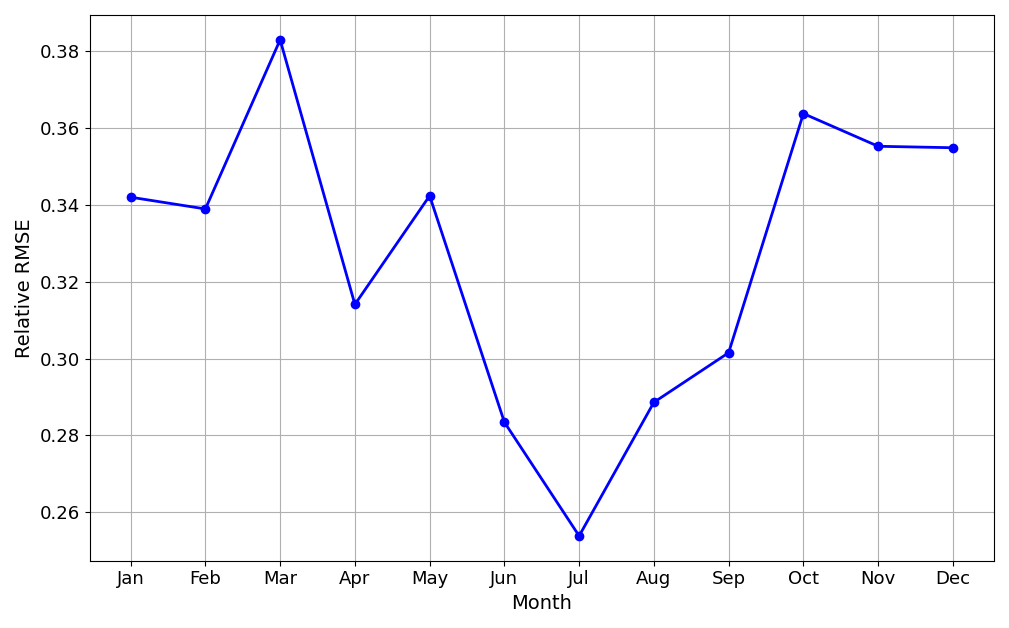}
    \caption{Monthly relative RMSE of the fine-tuned $\mathcal{M}'$ model, averaged over 21 Corsican stations and all forecast horizons (1–6 hours). Higher values indicate reduced model accuracy.}
    \label{fig:monthly_relative_rmse}
 \vspace*{-0.2cm}
\end{figure}

It is relevant to assess whether model performance exhibits seasonal variability. To this end, we computed the monthly relative RMSE (denoted as RRMSE and defined as RMSE divided by mean wind speed) of the fine-tuned $\mathcal{M}'$ model, averaged over all 21 Corsican stations and forecast horizons (1–6 hours). Figure~\ref{fig:monthly_relative_rmse} displays the results. It can be observed that the relative RMSE is highest during the extended winter October–March, indicating reduced model accuracy during these months. This period is characterized by more dynamic and variable wind regimes, which may pose greater challenges for prediction. In contrast, the lowest relative RMSE values are observed during June-August, suggesting better model performance under more stable summer conditions with notably well established breeze regimes.

\begin{table*}[h!]
	\centering
	\caption{Comparative RMSE for models $\cM'$ and $\cal N$ for horizons 1,3,6 hours.}
	\label{tab:rmse_improvement}
	\small
	\begin{tabular}{lS[table-format=1.0]S[table-format=1.2]S[table-format=1.2]S[table-format=2.1]}
		\toprule
		Site    & {Horizon (h)} & {Model ${\cal N}$ ($m s^{-1}$)} & {Model $\cM'$ ($m s^{-1}$)} & {Improvement (\%)} \\
		\midrule
		\multirow{3}{*}{Ajaccio} & 1 & 1.01 & 0.95 & 5.9 \\
		& 3 & 1.15 & 1.00 & 13.0 \\
		& 6 & 1.26 & 1.07 & 15.1 \\
		\midrule
		\multirow{3}{*}{Lucciana} & 1 & 1.08 & 1.03 & 4.6 \\
		& 3 & 1.24 & 1.16 & 6.5 \\
		& 6 & 1.35 & 1.22 & 9.6 \\
		\midrule
		\multirow{3}{*}{Figari}  & 1 & 1.21 & 1.15 & 5.0 \\
		& 3 & 1.41 & 1.27 & 9.9 \\
		& 6 & 1.56 & 1.32 & 15.4 \\
		\midrule
		\multirow{3}{*}{Renno}   & 1 & 0.97 & 0.92 & 5.2 \\
		& 3 & 1.10 & 1.00 & 9.1 \\
		& 6 & 1.14 & 1.05 & 7.5 \\
		\bottomrule
	\end{tabular}
\end{table*}

Finally, Table \ref{tab:rmse_improvement} compares the performance of the fine-tuned $\cM'$ against a neural network model introduced in \citet{BaileMuzy2023} which was trained using a dataset spanning a longer period (2011-2020) than that of MeteoNet.
This model has an architecture which is 
similar to the $\cM_{GS}$ model described in Sec. \ref{sec:model_arch} and takes, as input data, the last few hours observations from the target station and its 15 neighboring stations. This baseline model will be referred to as $\cal N$. Due to its relatively simple structure, $\cal N$ is  trained and optimized separately for each station. The reported RMSE values for both models are computed over the same evaluation set to ensure a fair comparison. The results show that, across all sites and forecast horizons, $\cM'$ consistently outperforms $\cal N$, despite being trained on a significantly shorter time period. Furthermore, as expected, the performance gap tends to widen with increasing forecast horizons. This highlights the advantage of $\cM'$, which integrates numerical weather prediction (NWP) data, over $\cal N$, which relies solely on past observed data. It also shows the importance of fine-tuning, particularly when training is conducted on a dataset that is too limited to fully optimize the model for local conditions.

\subsubsection{Predicting strong wind occurrences }
\label{sec:ext_winds}
\begin{figure}[h!]
	\centering
	\includegraphics[width=1.1\linewidth]{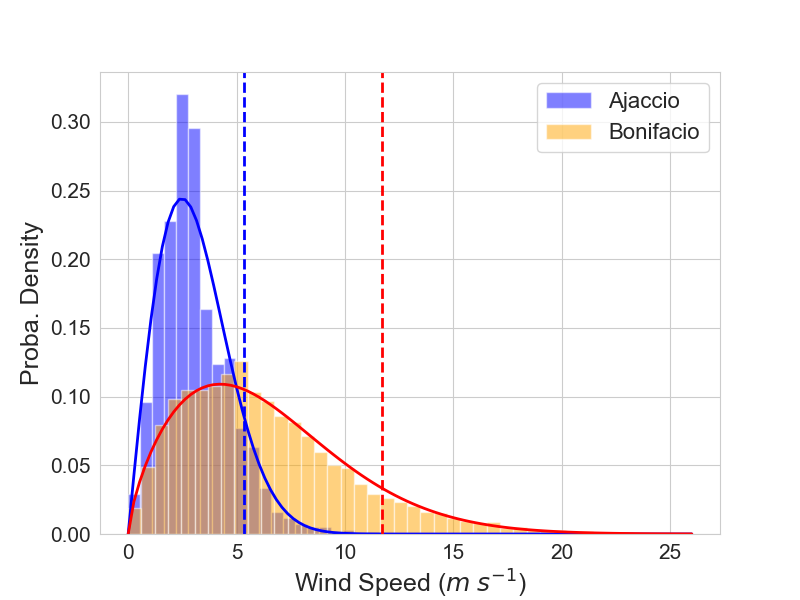}
	\caption{Wind speed distributions observed at the Ajaccio (blue) and Bonifacio (orange) ground stations. Solid lines represent the corresponding Weibull distribution fits. Dashed lines indicate the 90 \% quantiles for both locations. Wind speed units are $m s^{-1}$.}
	\label{Fig:ext_example}
\vspace*{-0.2cm}
\end{figure}

One possible appealing application of finely-tuned model $\cM'$, which is both local and probabilistic, is to predict the occurrence of local strong wind regimes. 
Indeed, the notion of strong wind velocity is very dependent on the site. In Fig. \ref{Fig:ext_example}, the observed wind velocity distributions in Ajaccio and Bonifacio  are illustrated. We see that
the wind distribution in Ajaccio is concentrated around low wind speeds, whereas in Bonifacio, the distribution is broader, indicating a higher likelihood of strong wind events. This reflects the distinct wind climates of the two locations: Bonifacio experiences significantly stronger and more turbulent winds. This is primary due to its exposure in the Strait of Bonifacio that channels and accelerates winds, particularly when they are northwesterly like Mistral.
The high limestone cliffs and surrounding hills also induce 
sudden wind gusts and direction shifts, making the wind highly variable near the coast. In contrast, Ajaccio is more sheltered within its gulf, resulting in generally milder winds, with only occasional strong gusts from dominant regional wind patterns (notably during Mistral and Libeccio episodes in autumn and winter). The concept of ``strong winds'' can therefore vary significantly depending on the location. 
As explained in Sec. \ref{sec:ext_metrics},
a practical approach to defining strong winds is to use a threshold based on the specific wind speed distribution of each site $s$.
This approach relates to an estimate of the potential wind impacts accounting for the adaptation of local infrastructure \citep{klawa_model_2003}.
The dashed lines in Fig. \ref{Fig:ext_example} represent the 90 \% quantile values for both locations, meaning that only 10 \% of wind occurrences exceed these thresholds at their respective sites.

Within this context, predicting strong wind events at a given location amounts to forecasting the occurrence of threshold exceedances corresponding a high quantile relative to the site's wind climatology.
In Sec. \ref{sec:ext_metrics}, such a quantity is denoted as 
$I_{s,d,t}(Q_C^s(p))$  (Eq. \eqref{eq:def_quantile}) with a probability level $p$ close to $p=1$.

In order to assess the quality of the prediction, we use PSS as discussed in Sec. \ref{sec:ext_metrics}  (Eq. \eqref{eq:PSS}). One of its main advantages is that it is unaffected by class imbalance, making it particularly useful in forecasting of strong wind episodes.

\begin{figure}[h!]
	\centering
	\includegraphics[width=0.9\linewidth]{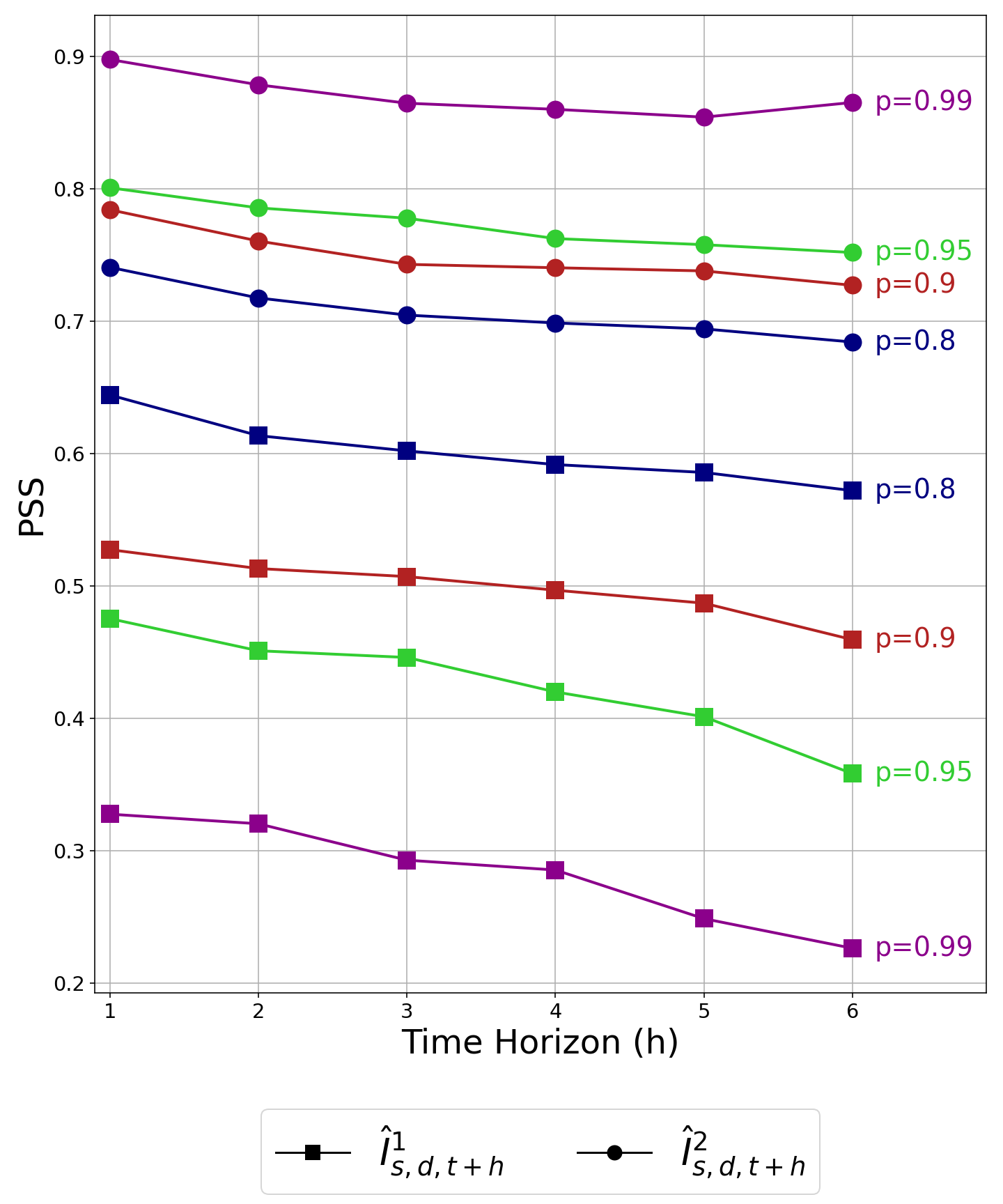}
	
	\caption{Predicting strong event occurrences: Symbols ($\blacksquare$) represent deterministic predictions (based on mean value) while symbols ({\Large $\! \! \bullet$}) represent probabilistic based prediction (based on quantile value). For each type of prediction, 4 curves are represented corresponding to a probability level $p=0.8,0.9,0.95,0.99$ that is used to define $I_{s,d,t}$ according to Eqs. \eqref{eq:pred_mean} and \eqref{eq:pred_quant}.}
	\label{Fig:ext1}
\vspace*{-0.2cm}
\end{figure}

\begin{figure*}[htbp]
	\centering
		\includegraphics[width=0.8\linewidth]{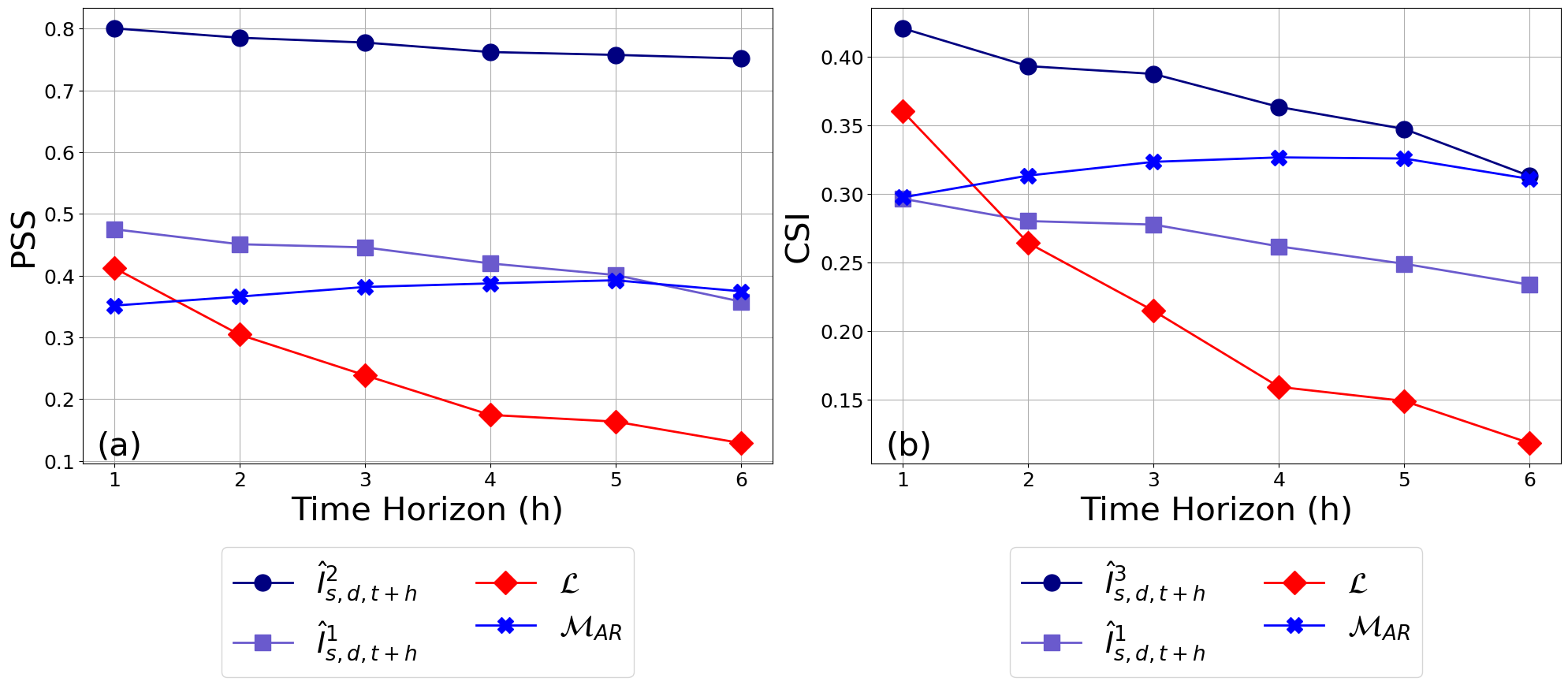}
	\vspace*{-0.5cm}
	\caption{(a) Predicting strong event occurrences: comparisons of three models for predicting the exceedance of the 95 \% quantile of each site. (\textcolor{blue!45!black}{ $\bullet$}) symbols represent the fine-tuned $\cM'$ model with predictor from Eq.~\eqref{eq:pred_quant}, $(\textcolor{violet}{\blacksquare})$ symbols represent the fine-tuned $\cM'$ model with predictor from Eq.~\eqref{eq:pred_mean}, (\textcolor{blue}{$\protect \fatx$}) symbols represent the prediction from $\cM_{AR}$ model and the (\textcolor{red}{$\blacklozenge$}) represent the performance of the linear prediction, where a linear regression over past observed wind speeds at the station and its surrounding stations is optimized, individually for each station, in order to minimize MSE. (b) The same models are compared in terms of CSI where the level $p'=0.33$ is used (Eq.~\eqref{eq:pred_quant_CSI}).}
	\label{Fig:ext2}
\vspace*{-0.2cm}
\end{figure*}

In the case of a single point prediction at site $s$, day $d$ and time $t$ the corresponding estimator is denoted as $\hI^1_{s,d,t+h}$ (Eq. \eqref{eq:pred_mean}) while estimator $\hI^2_{s,d,t+h}$ (Eq. \eqref{eq:pred_quant}) is built from a probabilistic output of model $\cM'$ in order to optimize PSS .
 In Fig. \ref{Fig:ext1} we compare the PSS value, averaged for all 21 Corsican stations,  as obtained on the test and validation periods, for the fine-tuned $\cM'$ model when one use estimators $\hI^1_{s,d,t+h}$ and $\hI^2_{s,d,t+h}$.
 Comparison is performed for all horizons $h = 1,2,3,4,5,6$ hours and threshold values corresponding to probability levels $p=0.8,0.9,0.95,0.99$ station climatology quantiles. The PSS performance is averaged across both the validation and test periods to provide more representative samples of extreme events, which becomes increasingly important as the quantile considered increases.
 We clearly observe that $I^2_{s,d,t+h}$, the prediction obtained by leveraging probabilistic forecasting to estimate conditional quantiles, achieves significantly better performance, particularly for the extreme quantiles.
 As the threshold value \( Q_C^{s}(p) \) increases when $p$ rises from 0.8 to 0.99, the predictive performance 
 of $\hI^1_{s,d,t+h}$ (based on the conditional mean \( \hV_{t+h}^s \), see Eq. \eqref{eq:pred_mean}) deteriorates sharply, whereas the performance of $\hI^2_{s,d,t+h}$ (based on the conditional quantile prediction \( \widehat{Q}^{s}_{t+h}(p) \), see Eqs \eqref{eq:def_cond_quantile} and \eqref{eq:pred_quant}), improves. In other words, probabilistic forecasting effectively captures extreme wind speed occurrences, outperforming the standard mean squared error (MSE)-based predictor.

Finally, the performance of the fine-tuned probabilistic model $\cM'$ for predicting strong wind speeds is assessed using the PSS and CSI metrics, with estimators defined in Eqs. \eqref{eq:pred_quant} and \eqref{eq:pred_quant_CSI} respectively. They are compared to $\hI^1$ based on a single-point prediction as defined in \eqref{eq:pred_mean} when using (i) the finely tuned $\cM'$ conditional mean as predictor, (ii) the prediction of  the linear model $\cal L$ introduced in Sec. \ref{sec:baselines} and optimized for each station and (iii) the prediction of the model $\cM_{AR}$ defined in Sec. \ref{sec:model_arch}. The results for all sites in Corsica over both validation and test intervals are presented for $p=0.95$ in Fig. \ref{Fig:ext2}(a) for the PSS and \ref{Fig:ext2}(b) for the CSI metrics. 
Regarding the PSS, as already found in Fig. \ref{Fig:ext1}, Fig. \ref{Fig:ext2}(a) shows that the ability of $\hI^1$  built using all the deterministic baseline models to predict strong wind events is significantly lower than that of the $\hI^2$ estimator obtained with the $\cM'$ model.  Incidentally, the model $\cal L$ outperforms $\cM_{AR}$ at short time horizons. However, its performance declines rapidly as the forecast horizon increases, which contrasts with the $\cM'$ and $\cM_{AR}$ models that both maintain stable predictive skill over time. Similar results are found in Fig. \ref{Fig:ext2}(b) for the CSI, although the benefits of using $\hI^3$ are less significant, particularly at largest horizons. This may be due to the fact that, as CSI becomes significantly smaller than 0.5, the chosen threshold corresponding to $p'=0.33$ becomes too large as respect to the optimal threshold. 

\section{Discussion about some limitations and specific issues.}
\label{sec:discussion}

\subsection{Temporal and spatial limitations}
It should be noted that the results of this study are based on a limited three-year temporal period (2016-2018).
Such a period is clearly  too short to capture all wind regimes, extreme occurrences or  interannual variabilities. However, we can first emphasize that the objective in this study is to design and evaluate a short-term wind speed forecasting method, not to assess the long-term wind resource potential or climatological characteristics of specific sites. In contrast to wind resource assessment studies—where long historical datasets are essential to estimate inter-annual variability and long-term extremes—this work focuses on optimizing a predictive model that delivers accurate short-term probabilistic forecasts using available real-time information. From this perspective, the key concern is not the long-term frequency of extremes, but rather the model’s ability to generalize to a broad range of conditions within the training and evaluation window.
Second, the proposed neural network architecture is trained to optimize full accuracy across the entire distribution of wind speeds. Notably in its probabilistic version that performs better for strong events, it leverages all available data to learn the parameters of the forecast distribution. Unlike models designed specifically to predict rare or extreme events—often relying on tail-focused objectives or imbalance-aware training strategies—the approach proposed in this work does not require over-representation of extremes in the training data. Instead, it is designed to perform well across the full spectrum of possible outcomes, including high wind speeds when they appear. Last, the extremes we consider in the evaluation are defined with respect to empirical quantiles computed from the 3-year dataset used in the study. That is, we use threshold exceedances defined within the climatology of the available period (e.g., 95th percentile of observed wind speeds at each site). By construction, such thresholds ensure that extreme events, as defined in this study, are present in the dataset and therefore adequately represented in the training and testing phases. This allows us to meaningfully assess the model's performance under “extreme” conditions, even within a 3-year window.
However, extending the evaluation to a longer historical period would help further validate the robustness of the method under broader climatological variability. 

Regarding the spatial scope of this study, the results presented in Fig.~\ref{fig:RMSEimpSE} indicate that the RMSE gains obtained when fine-tuning the globally trained model are consistent across the various subregions of the southeastern MeteoNet domain, and comparable to those observed in Corsica. This outcome suggests that the proposed two-step training and fine-tuning strategy is transferable and can be successfully replicated within other parts of the same broad climatic zone. Nevertheless, we recognize that regional variability, particularly across distinct climatic regimes, may influence the model’s performance. Extending the evaluation to the northwestern (Atlantic) regions of MeteoNet, characterized by meteorological conditions markedly different from those of the Mediterranean, represents an important direction for future work.  

\subsection{Considerations for wind energy applications}

In the context of wind energy forecasting, it is important to recognize that standard meteorological stations typically measure wind speed at heights around 10~m, whereas wind turbine hub heights are commonly between 80 and 140~m. This vertical discrepancy presents a known challenge in operational forecasting. While the models proposed in this paper are trained using near-surface observations and NWP fields, their relevance to wind energy applications is supported by established extrapolation techniques. These include empirical methods such as the power law and logarithmic wind profiles~\citep{Gualtieri2015, Lopez2022}, as well as site-specific shear estimates from the integration, with long-term mast observations, of reanalysis data~\citep{Gualtieri21, Yang2024} or  short-term LiDAR measurements \citep{Floors11}.  Additionally machine learning approaches capable of capturing non-linear atmospheric dynamics~\citep{Gualtieri2019} have been successfully used to model vertical wind profiles and extend near-surface forecasts to hub height. While these methods enhance the practical utility of near-surface forecasts, further validation based on actual measurements at hub height remains necessary~\citep{Drechsel12} and is identified as an interesting prospect for future work.

One can mention another limitation of the present study: although its evaluation focuses on point-wise accuracy metrics such as measured by  RMSE, it is important to note that operational wind energy applications are highly sensitive not only to the mean forecast accuracy but also to the temporal structure and variability of wind speed.  In this context, a detailed analysis of the spectral content and temporal correlation structure of the predicted wind signals is crucial for properly characterizing ramping behavior and short-term fluctuations.
As a perspective, future work will be dedicated to computing variance-based metrics (e.g., standard deviation ratio, normalized RMSE) and conducting spectral error analyses, in order to assess how well the models $\mathcal{M}$ and $\mathcal{M}'$ capture the multi-scale temporal variability of wind fluctuations.

\subsection{Operational feasibility}
\begin{table*}[tbh!]
	\centering
	\caption{Summary of Hourly Prediction Chain (with 6-hour NWP refresh)}
	\label{tab:chain}
	\begin{tabular}{@{}llp{3.3cm}l@{}}
		\toprule
		\textbf{Step} & \textbf{Frequency} & \textbf{Description} & \textbf{Execution Time} \\
		\midrule
		NWP forecast availability & Every 6 h & From meteorological center & 1–3 h \\
		GRIB file download        & Every 6 h & Transfer NWP forecasts locally        & $\sim$1 min \\
		Ground station data update & Every hour & Load observations from file or API      & $\sim$1 min \\
		Feature preparation       & Every hour & Normalize and assemble input tensors (NWP + station) & $\sim$1 min \\
		NN model inference        & Every hour & Forecast wind speed for 1–6~h lead times         & $\sim$5 s \\
		\bottomrule
	\end{tabular}
\end{table*}

To assess the practical applicability of the wind speed forecasting approach developed above, one can outline a theoretical operational prediction chain aligned with current production cycles of NWP systems such as AROME and ARPEGE. These models typically provide forecasts four times daily (i.e., initialized at 00, 06, 12 and 18 UTC), which become available approximately 1–3 hours later. Once available, GRIB files are downloaded and parsed, and the relevant fields are stored locally for use throughout the subsequent 6-hour cycle.

In the framework proposed by this study, the hybrid model $\cM$ or $\cM'$—designed to produce forecasts from 1 to 6 hours ahead—is executed hourly. At each prediction time, the latest ground station measurements are retrieved, and input tensors are assembled from the stored NWP fields and recent observations from nearby ground stations. The inference step is computationally lightweight: on a standard laptop, the entire prediction cycle—including data loading, preprocessing, and model inference—takes less than 5 seconds.

Table~\ref{tab:chain} summarizes the timing and sequence of operations in the hourly prediction loop. This architecture ensures low-latency and high-frequency forecasting, compatible with real-time operational constraints. By leveraging pre-stored NWP fields and hourly updates from ground stations, the system maintains responsiveness and computational efficiency, making it suitable for applications such as wind energy management, local warnings, and short-term response.

\section{Summary and prospects}
\label{sec:Conclusion}

This study introduces a hybrid neural network model for short-term (1-6 hours ahead) surface wind speed forecasting, combining Numerical Weather Prediction (NWP) data with observational data. The model leverages the MeteoNet dataset, which includes global ARPEGE and high-resolution AROME NWP data and observations from ground weather stations in Southeast France, with a focus on the Mediterranean island of Corsica. 
The proposed model, denoted as $\mathcal{M}$ (or $\mathcal{M}'$ for its probabilistic version), integrates these three sources of input data within a neural network architecture that combines LSTM units to process temporal information and convolutional layers to capture spatial patterns. $\cM$ is trained to minimize the mean squared error (MSE) for deterministic forecasts while, for probabilistic forecasts, $\cM'$  predicts the parameters of a Multifractal-Rice (M-Rice) distribution, achieved by minimizing the associated negative log-likelihood.

The hybrid model proposed in this work significantly outperforms baseline models, including raw NWP predictions (AROME and ARPEGE), persistence models, and linear regression models, across all forecast horizons (1-6 hours). It also performs better than a neural network specifically optimized to handle recent observations at a given ground station and its neighboring sites as proposed in \citet{BaileMuzy2023}. When analyzing feature importance, it can be observed that wind components are the most important features for prediction accuracy. AROME features remain consistently important across all horizons, while ARPEGE features gain importance at longer horizons. Direct observations from ground stations are more useful for forecasts at shortest horizons.
The probabilistic version of the model appears to slightly outperforms the deterministic version in terms of RMSE. Fine-tuning the model for specific stations further improves forecasting accuracy, particularly for smallest time horizons. This approach that leverages the optimal weights obtained from the global training phase mitigates overfitting when one wants to adapt the model to a specific location. 

In the context of extreme wind prediction, defined as the occurrence of wind speeds exceeding high quantiles (e.g., 90 \%, 95\%, 99\%) of site-specific distributions, our results demonstrate that the probabilistic model significantly outperforms linear regression models and post-processed AROME predictions, especially at longer forecasting horizons. It is also shown that our approach yields much more accurate predictions compared to its deterministic counterpart, which is based solely on the expected mean value. Moreover, the model’s performance improves as the quantile increases. This observation will have to be further investigated using a larger dataset, as it suggests that, in terms of PSS score, conditions leading to very strong wind events may be easier to anticipate than those associated with milder wind speeds.

In conclusion, the hybrid neural network model demonstrates significant improvements in short-term wind speed forecasting by integrating NWP data with ground station observations. Its ability to provide both deterministic and probabilistic forecasts makes it a valuable tool for applications in renewable energy, public safety, and operational decision-making. Many avenues can be explored to build upon this work.
First, to better understand the proposed model limitations and guide targeted improvements, in continuation of the preliminary findings of Secs \ref{sec:perfs_vs_ws} and \ref{sec:seas_perf}, a promising direction for future work is to investigate the specific conditions that influence the magnitude of prediction errors. Such an analysis could include evaluating conditional errors stratified by various key factors such as wind regime, terrain complexity, or seasonal variability as first experimented in  Sec.~\ref{sec:seas_perf}. Secondly, an important perspective for future work is to focus on the probabilistic aspects of the model to improve its reliability in terms of probabilistic metrics as logScore or CRPS \citep{BaggioMuzy2024} and to compare it with other related approaches (for instance, non-parametric
models based on normalizing flows \citep{Marcille24}).
Additionally, our predictions could be improved by leveraging ensemble NWP forecasting, as pioneered by \citet{rasp2018}. This can  potentially lead to better risk assessment for extreme events \citep{Primo24}. 
Another promising direction involves incorporating additional data sources such as orography or surface roughness, that have been shown to influence the reliability of wind speed predictions \citep{rasp2018,Vel21}. The present study focuses on  short-term forecasts (1–6 hours ahead) due to the use of the MeteoNet dataset. However, our model is inherently capable of handling longer forecast horizons so  we plan to extend this approach to full intra-day (1-24 h) and day-ahead ($>$ 24 h) forecasts, by leveraging datasets that provide more frequent NWP updates over longer time ranges. Finally, advancements in deep learning architectures present promising opportunities for further refinement. While LSTMs and CNNs have demonstrated strong performance, recent advances in Transformer-based models have shown great potential for capturing long-range dependencies in spatiotemporal data (see, e.g., \citet{transf23}).


%



\appendix


\section{The M-Rice probability distribution}
\label{app:M-Rice}
The Rice probability distribution corresponds to the amplitude (i.e. the norm) of a two dimensional random vector which components are 2 independent Gaussian random variables of mean $\mu_1$ and $\mu_2$ and of same variance $\sigma^2$. The corresponding probability density function, $f_R(z,\nu,\sigma)$ is the following \citep{Rice}:
\begin{equation}
\label{def:rice}
  f_{\text{R}}(y | \nu,\sigma) = \frac{y}{\sigma^2}e^{-\frac{y^2+\nu^2}{2 \sigma^2}} \; I_0 \Big(\frac{y \nu}{\sigma^2} \Big)
\end{equation}
where $\nu = \sqrt{\mu_1^2 + \mu_2^2}$ and $I_0(z)$ is the order zero modified Bessel function of the first kind.

Rice distribution has been extended by \citet{BaMuPo11} to ``Multifractal Rice" (M-Rice) distribution that accounts for the situation when, 
as observed in turbulence models, the variance $\sigma^2$ is itself stochastic with a log-normal distribution.  
The M-Rice distribution involves 3 parameters, namely the two Rice parameters  coming from from Gaussian law $\nu$ and $\sigma^2$ and a supplementary parameter, denoted as $\lambda^2$ associated with the variance of the log-normal law. This parameter is referred to, in the literature on turbulence, as the ``intermittency coefficient'' \citep{Fri95}. The M-Rice probability density function (PDF) is then:
\begin{eqnarray}
  \nonumber
	&& f_{\text{MR}}(y | \nu, \sigma, \lambda^2 ) =  \frac{1}{\sqrt{2 \pi \lambda^2}}  \int e^{-\frac{\omega^2}{2 \lambda^2}} f_{\text{R}} (y | \nu, \sigma e^{\omega}) \; d \omega  \\
\label{def:M-Rice2}
	& & =  \!  \frac{1}{\sqrt{2 \pi \lambda^2}}  \! \! \int \!\!\!\! e^{-\frac{\omega^2}{2 \lambda^2}} \! \!   \frac{y}{e^{2\omega}\sigma^2} e^{-\frac{y^2+\nu^2}{2 e^{2 \omega} \sigma^2}} I_0(\frac{y\nu} {e^{2 \omega} \sigma^2})  d \omega \; . \label{def:M-Rice}
\end{eqnarray}

As advocated in \citet{BaggioMuzy2024}, this last formula is evaluated using the following Gauss-Hermite quadrature:
\begin{equation}
	\int _{-\infty }^{+\infty }e^{-y^{2}}f(y)\,dy \approx \sum _{i=1}^{n}w_{i}f(y_{i})
\end{equation}
were $n \geq 1$, the abscissa $\{y_i\}_{i=1,\ldots,n}$ correspond to the roots of the order $n$-th Hermite polynomial $H_n(y)$ and the weights $\{w_i\}_{i=1,\ldots,n}$ are:  
$$
w_{i}={\frac {2^{n-1}n!{\sqrt {\pi }}}{n^{2}[H_{n-1}(y_{i})]^{2}}}
$$
with $n=11$ that is sufficient for the purpose of this paper. 

The M-Rice cumulative distribution function (CDF) function, useful to estimate the quantiles, can be obtained using the same quadrature formula and replacing the Rice PDF expression in \eqref{def:M-Rice} by the Rice CDF. This is also valid for the computation of the mean value: since for a Rice law of parameter $\nu$ and $\sigma^2$, the mean value is $\mu_{\text{R}} = {\displaystyle \sigma {\sqrt {\frac{\pi}{2}}}\,\,L_{\frac{1}{2}}\left(-\frac{\nu^{2}}{2\sigma ^{2}}\right)}$,
where $L_{\frac{1}{2}}$ stands for the order $\frac{1}{2}$ Laguerre polynomial,
the mean value of a M-Rice distribution reads:
\begin{eqnarray}
	\nonumber
	\mu_{\text{MR}} (\nu, \sigma, \lambda^2 )	& = &\frac{\sigma}{ \sqrt{2}}  \int e^{-y^2} \, L_{\frac{1}{2}}\left(-\frac{e^{2 \sqrt{2} \lambda y} \nu^{2}}{2\sigma ^{2}}\right)  \; d y  \\
	& \simeq & \frac{\sigma}{ \sqrt{2}} \sum_{i=1}^n w_i \, L_{\frac{1}{2}}\left(-\frac{e^{2 \sqrt{2} \lambda y_i} \nu^{2}}{2\sigma ^{2}}\right)  \label{eq:mean_mrice} \; .
\end{eqnarray}

\section*{Open Research Section}

\subsection*{Data Availability}

The data used in this study originate from the MeteoNet database~\citep{meteonet_dataset}, originally developed by Météo-France. A reference version used in this work is available on the Harvard Dataverse at \url{https://doi.org/10.7910/DVN/NCKRZ2}.

\subsection*{Code Availability}
The source code used in this study, including data preprocessing functions, model implementations, analysis scripts, and a minimal example dataset, is publicly available under the MIT License at \url{https://doi.org/10.5281/zenodo.15222910} (see \citet{saphir_predict_software}).

\section*{Acknowledgments}
This work was supported by ANR grant SAPHIR project ANR-21-CE04-001403.
The authors would like to acknowledge Marie-Laure Nivet (SPE) for insightful discussions and valuable assistance on matters related to open science.




\end{document}